\documentclass[review]{elsarticle}
\usepackage[pdftex]{hyperref}
\usepackage{array}
\usepackage{wrapfig}
\usepackage{amsmath, amsthm, amsfonts}
\usepackage{graphicx}
\usepackage{mathtools}
\usepackage{url}
\usepackage{array}
\usepackage{multirow}
\usepackage{caption}
\usepackage{subcaption}
\usepackage{color,soul}
\usepackage{setspace}
\usepackage{mathtools}
\usepackage{float}
\usepackage{blkarray, bigstrut}
\usepackage{appendix}
\usepackage{enumerate}
\usepackage{hyperref}
\usepackage[shortlabels]{enumitem}
\usepackage[linesnumbered,ruled,vlined]{algorithm2e}
\newtheorem{Ex}{Example}
\newcommand{\blackslug}{\rule{2mm}{2mm}}
\begin{document}

\begin{frontmatter}
\title{Investigating Entropy for Extractive Document Summarization}

\author[mainaddress]{Alka Khurana\corref{mycorrespondingauthor}}
\date{}
\cortext[mycorrespondingauthor]{Corresponding author}
\ead{akhurana@cs.du.ac.in}

\author[mainaddress]{Vasudha Bhatnagar}
\ead{vbhatnagar@cs.du.ac.in}
\address[mainaddress]{Department of Computer Science, University of Delhi, Delhi 110007, India}

\begin{abstract} 
Automatic text summarization aims to cut down readers' time and cognitive effort by reducing  the content of a text document without compromising on its essence. Ergo, \textit{informativeness} is the prime attribute of document summary generated by an algorithm, and selecting sentences  that capture the essence of a document is the primary goal of extractive document summarization.

In this paper, we employ Shannon's entropy to capture \textit{informativeness} of sentences. We employ Non-negative Matrix Factorization (NMF) to reveal  probability  distributions for computing  entropy of  terms, topics, and sentences in latent space.  We present an information theoretic  interpretation of the computed entropy, which is the bedrock of the proposed \textit{E-Summ} algorithm, an  unsupervised method for  extractive document summarization. The algorithm   systematically applies information theoretic principle  for selecting informative sentences from important topics in the document. The proposed algorithm is generic and fast, and hence amenable to use for summarization of documents in real time. Furthermore, it is   domain-, collection-independent and agnostic to the language of the document. Benefiting from strictly positive NMF factor matrices, \textit{E-Summ} algorithm  is transparent and explainable too.  

We use standard ROUGE toolkit for performance evaluation of the proposed method on four well known public data-sets.  We also perform quantitative assessment of  \textit{E-Summ} summary quality by computing its semantic similarity w.r.t the original document. Our investigation reveals that though using NMF and  information theoretic approach for document summarization promises  efficient, explainable, and language independent text summarization, it needs to be bolstered to match the performance of deep neural methods.

\end{abstract}


\end{frontmatter}

%%%%%%%%%%%%%%%%%%%%%%%%%%%%%%%%%%%%%%%%%%%
%%%%%%%%%%  SECTION - 1 %%%%%%%%%%%%%%%%%%%
%%%%%%%%%%%%%%%%%%%%%%%%%%%%%%%%%%%%%%%%%%%

\section{Introduction}
\label{sec-introduction}
Exponential growth in textual information available online has spurred the need for automatic processing of   text  for various  tasks that humanity performs on computing devices. Automatic document summarization  is one such task with a compelling ability to combat the problem of information overload. Proposed more than six decades ago by \cite{luhn1958automatic}, modest progress in the area of automatic summarization is evident by moderate scores achieved by sophisticated deep neural methods  \citep{fang2017wordcorank, nallapati2017summarunner, dong2018banditsum, zhou2018neural, narayan2018ranking, yasunaga2019scisummnet, alguliyev2019cosum, dou2021gsum, MATCHSUM_zhong2020extractive, khurana2020nmfensembles}. Further, recent debate and consequent surge in study of evaluation metrics for automatic summaries is a clear and strong testimony to the considerable complexity of the task \citep{peyrard2019simple, peyrard2019studying, ermakova2019survey, bhandari2020metrics, vasilyev2020human, fabbri2020summeval, huang2020achieved, bhandari2020re-evaluation}.

Research in designing and advancing automatic document summarization methods is largely driven by the objective to perform well on a few standard data-sets \citep{peyrard2019simple}.  The author further argues that community has exerted over \textit{crafting} algorithms to improve performance evaluation scores on the benchmark  data-sets, thereby limiting  progress in the science of automatic extractive document summarization. Similar searching questions have been asked by \citet{huang2020achieved}, who design a multi-dimensional quality metric and quantify  major sources of errors on well-known  summarization models. The authors underscore faithfulness and factual-consistency of extractive
summaries compared to abstractive counterparts, based on the designed metric.

Existing extractive summarization methods rely  on the intuitive notions of \textit{non-redundancy, relevance,} and \textit{informativeness} as signals of appositeness   for inclusion  of sentences in summary. While non-redundancy and relevance have been modeled in several earlier works \citep{luo2010effectively, alguliev2011IntegerLinearProgramming, gupta2014text, TGRAPH2015, nallapati2016classify,  nallapati2017summarunner, best2017cist, alguliyev2019cosum, saini2019extractive}, the notion of informativeness of a sentence  is completely unattended. Despite sophisticated supervised and unsupervised techniques, existing approaches for document summarization are fundamentally devoid of any theory of \textit{importance}.  \cite{peyrard2019simple} contends that the notion  of  \textit{importance} unifies non-redundancy, relevance and informativeness, the summary attributes hitherto  addressed  by the research community in an adhoc manner. 

Emphasis on modeling the vague human intuition of \textit{importance} in complete absence of a theoretical model is the primary impediment in advancing the science  of automatic document summarization to  technology.  \cite{peyrard2019simple} proposes to mitigate the problem by gleaning  probability distribution of semantic units of the document and computing entropy, to encode \textit{non-redundancy, relevance}, and \textit{informativeness} of a semantic unit as a single attribute, i.e., \textit{importance}.  Admitting sentences and topics as crucial semantics units  for document summarization, we present an in-depth analysis of sentence and topic entropy in latent semantic space revealed by Non-negative Matrix Factorization (NMF). Based on entropy, we propose an effective  sentence  scoring function for generating  document summary. Our research contributions are listed below. 

% encoding

\begin{enumerate}
    \item We delve into the latent semantic space of the document exposed by NMF (Sec. \ref{sec-int-latent-space}). We compute probability distribution and  entropy of semantic units and present corresponding interpretations (Sec. \ref{sec-sentence-entropy}, \ref{sec-topic-entropy}). We deliberate over the complex interplay of topic and sentence entropy by studying its impact on summary quality and corroborate our observations with pertinent empirical analysis (Sec. \ref{sec-selecting-summ-sen}).
    \item  We propose \textit{E-Summ}, an unsupervised, generic, explainable,  and language agnostic algorithm  for extractive document summarization (Sec. \ref{sec-algorithm}). 
     
    \item We use four well-known public data-sets (Sec. \ref{sec-experimental-design}) and present comparative evaluation of \textit{E-Summ} with recent unsupervised and deep neural methods (Sec. \ref{sec-experimental-results}). We also establish domain and language independence of the proposed method in the same section.
    
    \item  We also evaluate algorithmic summary quality by computing its semantic similarity with the complete document and show that the reference summaries have relatively less semantic overlap compared to \textit{E-Summ} summaries (Sec. \ref{sec-sem-sim-results}).

    \item We observe that despite a sound theoretical foundation, \textit{E-Summ} is not able to match the ROUGE score of deep neural methods.  We discuss this observation in detail in Sec. \ref{sec-discussion}.
     
\end{enumerate}

%%%%%%%%%%%%%%%%%%%%%%%%%%%%%%%%%%%%%%%%%%%
%%%%%%%%%%  SECTION - 2 %%%%%%%%%%%%%%%%%%%
%%%%%%%%%%%%%%%%%%%%%%%%%%%%%%%%%%%%%%%%%%%

\section{Background and Related Work}
\label{sec-related-work}
In this section, we first describe the proposal forwarded by \cite{peyrard2019simple}, which inspires the current work. Next, we present a review of works that uses entropy for document summarization. A subsection on Non-negative Matrix Factorization (NMF) follows, which is the method used to divulge the latent semantic space of the document.

\subsection{Information theoretic approach for Document Summarization}
\label{sec-info-approach}
Peyrard assumes that a document $D$ can be represented as probability distribution $P_D$ over the semantic units, $\Omega = \{\omega_1, \omega_2, \ldots, \omega_m\}$, where $P_D(\omega_i)$ could be interpreted either as the frequency of  unit $\omega_i$ or its normalized contribution to the meaning of $D$. Terms, topics, frames,  embedding, etc., present as possible semantics units comprising the document. 

Relevance of the semantic units comprising the summary is critical for reducing the uncertainty about $D$. Low redundancy of semantic units in summary implies high coverage. High coverage in turn translates to high informativeness, which is modeled by entropy in a straightforward manner. Thus high entropy semantic units, when included in summary, automatically reduce redundancy and augment informativeness.

\cite{peyrard2019simple} claims that once the background knowledge ($K$) of the reader is modeled, information theoretic framework is powerful enough to create personalized summaries. Importance of the semantic units to be included in summary blends informativeness and relevance. Author characterizes the \textit{importance} of the semantic unit $\omega_i$ using function $f(d_i, k_i)$ where $d_i$, $k_i$ are respective probabilities of $\omega_i$ in document $D$ and background knowledge $K$. Here, $f(d_i, k_i)$ encodes importance of semantic unit $\omega_i$ and is required to satisfy the following conditions.
\begin{enumerate}[(i)]
    \item Informativeness: $\forall i \neq j$, if $d_i = d_j$ and $k_i > k_j$ then  $f(d_i, k_i) < f(d_j, k_j)$. This prefers inclusion of $\omega_j$ in summary, which is more informative for the user.
    
    \item Relevance: $\forall i \neq j$, if $d_i > d_j$ and $k_i = k_j$ then  $f(d_i, k_i) > f(d_j, k_j)$. This condition implies that out of two  semantic units - $\omega_i$ and $\omega_j$, occurring with equal probability in background knowledge ($K$) of the user, one that is more frequent in $D$ is relevant for the summary.

    \item Additivity: This condition combines informativeness and relevance, using Shannon's theory of information, defined as:
    $I(f(d_i, k_i)) \equiv \alpha I(d_i) + \beta I(k_i)$, where $\alpha, \beta$ are user-defined parameters representing strength of relevance and informativeness respectively.
    
    \item Normalization: $\sum_{i}f(d_i, k_i) = 1$, which ensures $f$ defines a valid probability distribution.
\end{enumerate}

Author establishes that a function satisfying above four requirements will have fixed form: 
$P_\frac{D}{K}(\omega_i) = \frac{1}{C} . \frac{d_i^\alpha}{k_i^\beta}$, where $C = \sum_{i}\frac{d_i^\alpha}{k_i^\beta}$, where $\alpha, \beta \in \mathbb R^+$(See \cite{peyrard2019simple} for proof). Therefore function $P_\frac{D}{K}$ implicitly codes relevance and informativeness.

\subsection{Entropy for Document Summarization}
\label{sec-ent-doc-summ}
Earlier works use \textit{entropy} for document summarization either as a method of ranking sentences or for evaluating summaries. \cite{kennedy2010entropy} use entropy to measure the quality of summary by calculating the amount of unique information captured in the summary. Words in the summary are mapped to concepts (topics) using Roget’s Thesaurus (\cite{jarmasz2004roget}), which contains approximately one thousand concepts with weight assigned to each concept. Normalizing the weights to obtain probability distribution, authors map summary words to the concepts and calculate entropy of the summary for quantitative assessment of its quality.

\cite{luo2010effectively} conjecture that sentence entropy proxies for \textit{coverage} of information by the sentence. Authors consider sentence as a vector of terms (content words) in the document and compute probability distribution of terms, which is used for calculating entropy of the sentence. Since high entropy of a sentence implies more coverage, the method has inherent bias towards long sentences, favoring their inclusion in summary. \cite{yadav2018new} gauge entropy in latent semantic space of the document by computing probability distribution of topics in sentences and vice-versa. They use Latent Semantic Analysis (LSA) to reveal the latent semantic space of the document. The interplay of topic and sentence entropy is used for selecting summary sentences. Since LSA factor matrices are used to compute the probability distribution of topics and sentences, authors are compelled to ignore negative terms in factor matrices, thereby losing substantial information. 

% Authors consider the sentence as a vector of terms (content words) in the document and compute the probability distribution of terms used to calculate the entropy of the sentence.  Since the high entropy implies more coverage, the method has an inherent bias towards long sentences, favoring their inclusion in summary. 

\subsection{NMF for Document Summarization}
\label{sec-nmf-doc-summ}
We use NMF to reveal the latent semantic space of the document. The preference for NMF over LSA is motivated by the presence of non-negative terms in NMF factor matrices, ensuing enhanced interpretability (\cite{lee1999nature}). Furthermore, the presence of non-negative terms overcomes the loss of information incurred by ignoring negative terms in LSA factor matrices. 

Non-negative Matrix Factorization (NMF) is a matrix decomposition method for approximating non-negative matrix $A$ ($\in$ ${\mathbb{R}}^{m\times n}$) as $A\approx WH$ in reduced space. Here, $W$ $\in$ ${\mathbb{R}}^{m\times r}$ and $H$ $\in$ ${\mathbb{R}}^{r\times n}$ are non-negative factor matrices and $r \ll \min\{m,n\}$. Starting with non-negative seed values for $W$ and $H$ matrices, NMF algorithm iteratively improves both factor matrices  to approximate $A$ by product of $W$, $H$ such that Frobenius norm ${\mid\mid A- WH \mid\mid}^2_F$ is minimized.

Considering document $D$ to be summarized as a sequence of $n$ sentences ($D=(S_{1}, S_{2}, \dots, S_{n})$), $D$ is represented as term-sentence matrix $A$  where columns of $A$ correspond to sentences and rows represent terms ($T = \{ t_{1}, t_{2}, \dots, t_{m} \}$). Accordingly, $A$ is an $m \times n$ matrix where element $a_{ij}$ in $A$ denotes occurrence of term $t_{i}$ in sentence $S_{j}$.

NMF decomposition of $A$ reveals latent semantic space of the document via. two non-negative factor matrices $W$ and $H$. Here, $W$ is $m \times r$ term-topic (feature) matrix and $H$ is $r \times n$ topic-sentence (co-efficient) matrix. Columns of matrix $W$ correspond to $r$ latent topics represented as ($\tau_{1}, \tau_{2} \ldots \tau_{r}$). Each element $w_{ij}$ in $W$ gives the strength of term $t_{i}$ in topic $\tau_{j}$ whereas element $h_{ij}$ in $H$ specifies the strength of topic $\tau_{i}$ in sentence $S_{j}$ and vice-versa. Both factor matrices $W$, $H$ in latent space and input matrix $A$ can be efficiently exploited for document summarization. 

\cite{lee2009automatic} propose NMF based unsupervised extractive document summarization method. The authors score the sentences by computing Generic Relevance Score (GRS) for each sentence. More recently, \cite{khurana2019extractive} propose NMF based methods for extractive document summarization. The methods use term-oriented and topic-oriented approach for scoring sentences in the document.

%%%%%%%%%%%%%%%%%%%%%%%%%%%%%%%%%%%%%%%%%%%
%%%%%%%%%%  SECTION - 3 %%%%%%%%%%%%%%%%%%%
%%%%%%%%%%%%%%%%%%%%%%%%%%%%%%%%%%%%%%%%%%%

\section{Interpreting  NMF Factor Matrices}
\label{sec-int-latent-space}
Every well written document has a theme, and  sentences in the document carry information about  topics within the scope of the theme. It is, therefore,  reasonable to posit that sentences and topics are two prime carriers of information (semantic units) contained in the text document. Extractive summary aims to extract \textit{informative} sentences that communicate \textit{important} topics in the document. 

Information in a document is woven as sentences by the terms (content words) glued by stop-words (non-content words). Further, since the information in the document is embodied by  both sentences and topics,  both sentence and topic entropy are potent  to gauge \textit{informativeness} of the document. Naturally, the two are intricately related by \textit{terms}, which are the common  foundational semantic units.  In this and the next section, we delve into the computation, interpretation, and the interplay of sentence  and topic entropy for generating informative summaries. 

%Further, since both sentences and topics embody the information in the document,  both sentence and topic entropy are potent to gauge \textit{informativeness} of the document.

\begin{figure}[h!]
\begin{minipage}{0.60\textwidth}
\includegraphics[width=\textwidth,height=0.40\textwidth]{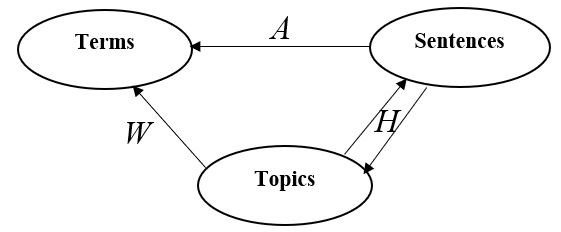}
\end{minipage}
\begin{minipage}{0.35\textwidth}
\caption{Relationship between terms, sentences and topics (in latent space) after NMF decomposition. $A$:term-sentence matrix, $W$:term-topic matrix, $H$:topic-sentence matrix}
\label{fig-term-topic-sent}
\end{minipage}
\end{figure}

\textit{Terms, sentences,} and \textit{topics} (in latent space), the three semantic units in a document, are inter-related, as shown in  Fig. \ref{fig-term-topic-sent}. Arrows denote the relation \textit{``consists-of"}, i.e.   sentences \textit{consists-of} terms, as do topics (in latent space). Sentences \textit{consists-of} topics (in latent space), while topics \textit{consists-of} sentences. Sentences and topics  are intricately intertwined as sentences discuss topics and topics are spread over sentences. Hence, the two have a symbiotic relationship.
 
Non-negative Matrix Factorization (NMF)  of term-sentence matrix ($A$) of a document reveals topics in the latent space through the feature matrix ($W$)  and co-efficient matrix ($H$). Quantitative relationships between the three semantic units of the document is encoded by  $A, W$ and $H$ matrices (Fig. \ref{fig-term-topic-sent}). Matrix $W$ quantifies contribution of terms in latent topics, while  $H$ quantifies a linear interdependent relationship between topics and sentences (Sec. \ref{sec-nmf-doc-summ}). 

\begin{Ex}
\label{exam-doc}
\begin{figure}[htbp!]
\centering
\caption{ Terms, Sentences and Topics in latent space for the sample document from DUC2002 data-set\textsuperscript{\ref{footnote-doc-no}}}
\label{fig-example-1}
\begin{subfigure}[b]{\textwidth}
\caption{Document Sentences}
\label{fig-example-1-a}
\noindent\fbox{%
    \parbox{\textwidth}{%
\scriptsize{\setstretch{1.0}{
\textbf{S1:} Emperor Hirohito has lost nearly 40 pounds in the two months since he fell ill and now weighs only 66 pounds, a palace official was quoted as saying Friday.

\textbf{S2:} The 87-year-old emperor's condition weakened Friday, but doctors speculated that a scab had stopped internal bleeding in his upper intestinal area, said Imperial Household Agency spokesman Kenji Maeda.

\textbf{S3:} Hirohito has suffered repeated bouts of bleeding but has not discharged blood in nine days.

\textbf{S4:} The emperor responds to queries from aides but is ``less talkative than before,'' Maeda told reporters.

\textbf{S5:} He added that doctors sometimes have trouble determining whether Hirohito is awake.

\textbf{S6:}  News reports quoted Maeda as saying Hirohito has shed much of his normal 104 pounds during his confinement and now weighs about 66 pounds.

\textbf{S7:} Hirohito had a temperature of 99.1 degrees Friday evening, up slightly from the morning.

\textbf{S8:} His normal temperature is about 95.9 degrees.

\textbf{S9:} Doctors have said he suffers from jaundice and anemia.

\textbf{S10:} Maeda said the emperor's heart and kidneys were functioning normally.

\textbf{S11:} Since Sept. 19, when he was confined to intensive care after vomiting blood, Hirohito has received 42 pints of blood in transfusions.

\textbf{S12:} Palace officials have not given details on Hirohito's illness, and they would not comment on Japanese news reports that the emperor has pancreatic cancer.

}}
}
}
\end{subfigure}

\begin{subfigure}[b]{\textwidth}
\caption{Latent topics consisting of terms extracted from matrix $W$ (strength of contribution omitted)}
\label{fig-example-1-b}
\noindent\fbox{%
    \parbox{\textwidth}{%
\scriptsize{
\textbf{Topic 1:}\{emperor,	hirohito, friday, yearold, emperors, condition, weakened, doctors, speculated, scab, stopped, internal,	bleeding, upper, intestinal	,area,	said, imperial,	household, agency, spokesman, kenji, maeda,	suffered,	repeated, bouts, discharged, nine, days, responds, queries,	aides, less, talkative,	told, reporters, added,	sometimes, trouble,	determining, whether, awake, shed,	much, normal, confinement, temperature,	degrees, evening, slightly,	morning	, suffers, jaundice, anemia, heart, kidneys, functioning,	normally\}

\textbf{Topic 2:}\{emperor,	hirohito, lost,	nearly,	pounds,	two, months, since,	fell, ill, weighs, palace, official, quoted, saying, friday, doctors, maeda,		responds, queries, aides, less,	talkative, told, reporters,	added, sometimes,	trouble, determining, whether, awake, news,	reports, shed, much, normal,	confinement, temperature, degrees, evening,	slightly, morning, heart, kidneys, functioning, normally\}

\textbf{Topic 3:}\{emperor,	palace,	maeda,responds,	queries, aides,	less, talkative, told, reporters, news, reports, shed, much, normal, confinement,	
heart, kidneys,	functioning, normally, officials, given, details, hirohitos,	illness, would,	comment, japanese, pancreatic, cancer\}

\textbf{Topic 4:}\{hirohito, since, friday,	doctors, bleeding, suffered,	repeated, bouts, discharged, blood,	nine, days, added,	sometimes, trouble,	determining, whether, awake, temperature, degrees, evening,	slightly, morning,	suffers, jaundice, anemia, sept, confined, intensive, care,	vomiting, received	pints, transfusions\}
}}}
\end{subfigure}

\begin{subfigure}[b]{\textwidth}
\caption{Latent topics discussed by sentences with their respective contributions in the topic (extracted from matrix $H$)}
\label{fig-example-1-c}
\noindent\fbox{%
    \parbox{\textwidth}{%
\scriptsize{
\textbf{Topic 1:}\{(S2, 2.504), (S3, 0.062), (S4, 0.189) (S5, 0.128), (S6, 0.035), (S7, 0.086), (S9, 0.330), (S10, 0.530)\}

\textbf{Topic 2:}\{(S1, 1.666), (S4, 0.137) (S5, 0.067), (S6, 1.096), (S7, 0.249), (S8, 0.087), (S10, 0.025)\}

\textbf{Topic 3:}\{(S4, 0.281), (S6, 0.100), (S10, 0.009), (S12, 1.956)\}

\textbf{Topic 4:}\{(S3, 1.032), (S5, 0.523), (S7, 0.357), (S8, 0.037) (S9, 0.019), (S11, 1.600)\}
}}}
\end{subfigure}

\ContinuedFloat

\begin{subfigure}[b]{\textwidth}
\caption{Sentences consisting of latent topics with their respective contribution in the sentence (extracted from matrix $H$)}
\label{fig-example-1-d}
\noindent\fbox{%
    \parbox{\textwidth}{%
\scriptsize{
\textbf{S1:}\{(Topic 2, 1.666)\}

\textbf{S2:}\{(Topic 1, 2.504)\}

\textbf{S3:}\{(Topic 1, 0.062), (Topic 4, 1.032)\}

\textbf{S4:}\{(Topic 1, 0.189), (Topic 2, 0.137), (Topic 3, 0.281)\}

\textbf{S5:}\{(Topic 1, 0.128), (Topic 2, 0.067), (Topic 4, 0.523)\}

\textbf{S6:}\{(Topic 1, 0.035), (Topic 2, 1.096), (Topic 3, 0.100)\}

\textbf{S7:}\{(Topic 1, 0.086), (Topic 2, 0.249), (Topic 4, 0.357)\}

\textbf{S8:}\{(Topic 2, 0.087), (Topic 4, 0.037)\}

\textbf{S9:}\{(Topic 1, 0.330), (Topic 4, 0.019)\}

\textbf{S10:}\{(Topic 1, 0.530), (Topic 2, 0.025), (Topic 3, 0.009)\}

\textbf{S11:}\{(Topic 4, 1.600)\}

\textbf{S12:}\{(Topic 3, 1.956)\}
}}}
\end{subfigure}
\end{figure}

Fig. \ref{fig-example-1} illustrates relationships between \textit{terms}, \textit{sentences}  and  \textit{topics} for an example document  from DUC2002 data-set\footnote{Document No. - AP881118-0104. The data-set is described in Sec. \ref{sec-experimental-design}. \label{footnote-doc-no}}(Fig. \ref{fig-example-1-a}). Application of community detection algorithm reveals four latent topics in the document (Sec. \ref{sec-topics}), which are exposed by non-negative matrix factorization method. Fig. \ref{fig-example-1-b} shows terms contributing in each latent topic in $W$ matrix. Note that some terms find mention in multiple topics, implying their semantic contribution in these topics. For example term ``Hirohito'' occurs in seven out of twelve sentences, contributing to three topics with different strengths.

%All terms with frequency one (occurring in one sentence) contribute to exactly one topic.  

%We identify the number of latent topics  by considering the document $D$ as weighted co-occurrence graph over terms. Application of Louvain algorithm on the graph reveals groups of related terms in the document, which correspond to latent topics in the document. The method for computing the number of latent topics is described in .

Semantic contribution of sentences to topics and vice-versa is revealed in Figs. \ref{fig-example-1-c} and \ref{fig-example-1-d} respectively. Fig. \ref{fig-example-1-c} shows the sentences contributing in each topic. Intensity of contribution of a sentence in a topic is mentioned in parentheses along with the sentence. For example, sentences $S4$, $S6$, $S10$ and  $S12$ contribute to $Topic\,3$ with strengths $0.281$, $0.100$, $0.009$, and $1.956$ respectively. Sentence $S12$ has highest contribution in $Topic\,1$ compared to other sentences. Fig. \ref{fig-example-1-d} shows sentences consisting of topics. Strength of a topic in a sentence  is mentioned in parentheses along with the contributing topic. For example, sentence $S3$ describes (consists-of)  $Topic\,1$ and $Topic\,4$, with the later having  higher strength.
\blackslug
\end{Ex}

%%%%%%%%%%%%%%%%%%%%%%%%%%%%%%%%%%%%%%%%%%%
%%%%%%%%%%  SECTION - 4 %%%%%%%%%%%%%%%%%%%
%%%%%%%%%%%%%%%%%%%%%%%%%%%%%%%%%%%%%%%%%%%

\section{Realizing Sentence and Topic Entropy}
\label{sec-realizing-info-summ}
Shannon entropy   quantifies the expected information conveyed by a random variable with specified probability distribution. It furnishes an  intuitive understanding of  the amount of uncertainty about an event associated with a given probability distribution. Hence it is a property of probability distribution of the event, with higher value of entropy implying  higher information content.

In order to compute sentence and topic entropy, it is imperative to mutate sentences and topics into their respective probabilistic versions. Since a sentence is composed of \textit{terms} and  \textit{  topics} in latent space simultaneously, probability distribution of sentences can be realized in two ways. Likewise, a topic is conjointly composed of \textit{terms} and \textit{sentences}, and its probability distribution can be realized using either \textit{terms} or \textit{sentences} (Refer to Fig. \ref{fig-term-topic-sent}).

\subsection{Computing Sentence Entropy}
\label{sec-sentence-entropy}
Entropy of a sentence quantifies the amount of information conveyed by the sentence. Sentences in a document are perceptible in both term space ($A$) as well as in latent topic space ($H$). Therefore, we compute sentence entropy in two ways - (i) in term space, we use probability distribution of terms over sentences, and (ii) in the space of latent topics, we use probability distribution of topics in sentences. 

\begin{enumerate}[(i)]
\item \textit{Sentence Entropy in term space}

Each column of term-sentence matrix $A$ is  the term vector for the corresponding sentence in the document. Probability ($p_{ij}$) of term $t_{i}$ in sentence $S_{j}$ is computed as $\frac{a_{ij}}{\sum\limits_{i=1}^{m} a_{ij}}$. Hence, entropy of sentence $S_{j}$ in term space is given by 
\begin{equation}
\label{eqn-sent-ent-A}
    \zeta^{t}(S_{j}) = -\sum\limits_{i=1}^m p_{ij} \log_2({p_{ij}})
\end{equation} 
Entropy of a sentence in term space is proportional to its length. Empirical correlation between sentence length and sentence entropy (calculated from matrix $A$) for four data-sets is shown in Row 1 of Table \ref{table-correl-sen-len-sen-entropy}. As conjectured by \cite{luo2010effectively}, there is a high positive correlation between sentence length and sentence entropy in term space.

\item \textit{Sentence Entropy in space of latent topics}

Column $H_{*j}$ of the co-efficient matrix $H$ quantifies the contribution of the sentence $S_{j}$ in $r$ latent topics. Let $H^{\prime}_{*j}$ be the corresponding normalized column, where element  $h^{\prime}_{ij}$ ( $= \frac{h_{ij}}{\sum\limits_{i=1}^{r} h_{ij}}$) is the probability of the latent topic $\tau_{i}$ in sentence $S_{j}$. Entropy of sentence $S_{j}$ in topic space is defined as 
\begin{equation}
\label{eqn-sent-ent-H}
    \zeta^{\tau}(S_{j}) = -\sum\limits_{i=1}^r h^{\prime}_{ij} \log_2({h^{\prime}_{ij}})
\end{equation}
High entropy sentences contribute in  multiple latent topics, and intuitively have larger coverage.   Row 2 of Table \ref{table-correl-sen-len-sen-entropy} shows low  correlation, slightly on the negative side, between sentence length and $\zeta^{\tau}(S)$. This implies that  sentence length is almost independent of entropy in latent topic space, i.e. longer sentences do not necessarily contribute to more topics. For example, sentence $S2$  in Fig. \ref{fig-example-1-a} is the longest sentence in the document, but in latent semantic space it contributes to only one topic, i.e. $Topic\,2$.

We also report statistical correlation between  $\zeta^{t}(S)$ and $\zeta^{\tau}(S)$ in Row 3 of Table \ref{table-correl-sen-len-sen-entropy}. Low correlation values for three data-sets reveal that sentence entropy in term space and that in latent topic space are nearly independent. Negative correlation between $\zeta^{t}(S)$ and $\zeta^{\tau}(S)$ for DUC2002 data-set can be explained as follows. Though a longer sentence has higher entropy in term space, there is no guarantee that it contributes to multiple latent topics. This possibly lowers the sentence entropy in latent topic space.  Fig. \ref{fig-example-1-a} shows  that $S2$ is the longest sentence in the document and has highest term space entropy. However, $S2$ contributes only in one latent topic and has zero topic space entropy. 

\begin{table*}
\centering
\caption{Pearson's Correlation Coefficient between \textbf{Row 1:}  sentence length and sentence entropy calculated in term space, \textbf{Row 2}:  sentence length and sentence entropy calculated from co-efficient ($H$) matrix, \textbf{Row 3:}  sentence entropies calculated from term-sentence matrix ($A$) and co-efficient ($H$) matrix, \textbf{Row 4}:  topic entropies calculated from NMF feature ($W$) and co-efficient ($H$) matrix}
\begin{tabular}{|p{4cm}|c|c|c|c|}
\hline
\textbf{Correlation Between}&\textbf{DUC2001} & \textbf{DUC2002}& \textbf{CNN}&\textbf{DailyMail} \\
\hline
Sentence length and $\zeta^{t}(S)$ & 0.811 & 0.856& 0.862 & 0.872\\
\hline
Sentence length and $\zeta^{\tau}(S)$ & -0.126 & -0.169  & -0.046 &-0.049\\
\hline
$\zeta^{t}(S)$ and $\zeta^{\tau}(S)$ & 0.012 & -0.025 & 0.056 &0.058\\
\hline
$\Psi^{t}(\tau)$ and $\Psi^{S}(\tau)$ &0.855 & 0.840  & 0.906  & 0.884\\
\hline
\end{tabular}
\label{table-correl-sen-len-sen-entropy}
\end{table*}

\end{enumerate}

\subsection{Computing Topic Entropy}
\label{sec-topic-entropy}
Topics in a document are distributed over terms as well as the  sentences in latent space. Probability distribution of topics over terms and sentences yields two articulations of topic entropy - (i) in the latent space of terms using feature matrix $W$, and (ii) in the space of sentences using co-efficient matrix $H$. 

\begin{enumerate}[(i)]
\item \textit{Topic Entropy in term space}

Column $W_{*i}$ quantifies the contribution of terms in corresponding latent topic. Let $\hat{W}_{*i}$ be the normalized column corresponding to $W_{*i}$, where $\hat{w}_{ji} = \frac{w_{ji}}{\sum\limits_{j=1}^{m} w_{ji}}$ is the probability of term $t_{j}$ in latent topic $\tau_{i}$. Entropy $\Psi^{t}(\tau_{i})$ of topic $\tau_{i}$ in term space is defined as 
\begin{equation}
\label{eqn-topic-ent-W}
    \Psi^{t}(\tau_{i}) = -\sum\limits_{j=1}^m \hat{w}_{ji} \log_2({\hat{w}_{ji}})
\end{equation}
High entropy of a topic in term space indicates that more terms comprise the topic with nearly equal probabilities. 

\item \textit{Topic Entropy in sentence space}

Row $H_{i*}$ in co-efficient matrix quantifies the contribution of $n$ sentences in the latent topic $\tau_{i}$. Let $\hat{H}_{i*}$ be the normalized row corresponding to $H_{i*}$, where $\hat{h}_{ij}$ is the probability of latent topic $\tau_{i}$ contributing in sentence $S_{j}$, computed as $\hat{h}_{ij} = \frac{h_{ij}}{\sum\limits_{j=1}^{n} h_{ij}}$.
Entropy of latent topic $\tau_{i}$ in sentence space is defined as 
\begin{equation}
\label{eqn-topic-ent-H}
    \Psi^{S}(\tau_{i}) = -\sum\limits_{j=1}^n \hat{h}_{ij} \log_2({\hat{h}_{ij}})
\end{equation}
High entropy of a topic in  space of sentences implies that the topic finds mention in multiple sentences.
\end{enumerate}

\subsection{Interpreting Sentence and Topic Entropy}
\label{sec-interpreting-SE-TE}
A ``good" extractive summary must comprise high quality semantic units to faithfully capture relevant information content from the document. Accordingly,  \textit{informative} summary can be realized  by including \textit{informative} sentences about \textit{important} topics discussed in the document.

Selecting sentences with high $\zeta^{t}(S)$ favours inclusion of longer sentences in summary. \cite{luo2010effectively} relate higher entropy of the sentence to higher coverage, thereby assuming that the summary with longer sentences covers more  aspects in the document than shorter ones. However, this assumption may not hold if the document is not well written. Often times, long sentences have diffused information either due to faulty construction or multiple ideas with little focus. Therefore, it is not prudent to employ $\zeta^{t}(S)$ as the sole criterion for scoring sentences. Recall that $\zeta^{\tau}(S)$ is length agnostic, and hence sentence entropy in topic space is better suited for measuring \textit{informativeness} of a sentence.  

High value of $\Psi^{t}(\tau)$ indicates higher information content of the latent topic in term space, which implies that the topic consists of more terms with nearly equal contributions. High value of $\Psi^{S}(\tau)$ signifies that the topic finds mention in many sentences.  We empirically analyze the correlation between $\Psi^{t}(\tau)$ and $\Psi^{S}(\tau)$.  Row 4 of Table \ref{table-correl-sen-len-sen-entropy} reveals high positive correlation between the two, suggesting that either of the two can be used interchangeably. We choose to use  $\Psi^{S}(\tau)$ as it is computationally efficient because of smaller size of $H$.

Even though high entropy signifies high informativeness,  it would be na\"ive to presume that selecting high entropy sentences describing high entropy topics is key to high quality summary.

\subsection{Selecting summary sentences}
\label{sec-selecting-summ-sen}
In this section, we examine the possible ways in which topic and sentence entropy ($\zeta^{\tau}(S)$ and $\Psi^{S}(\tau)$, respectively) can be meaningfully combined to extract \textit{informative} sentences from the document. Since low entropy implies less information, combination of low topic and low sentence entropy begets the negative effects of both, thereby degrading the quality of summary.  Selecting  sentences with high  entropy from low entropy topics too is not prudent either, since low entropy topic is discussed in fewer sentences and is probably not important. We investigate two promising combinations viz. \textit{High TE \& High SE} and \textit{High TE \& Low SE},  in order to gain better insight into the interplay between topic and sentence entropy. Macro-averaged ROUGE scores of summaries generated by these combinations is presented in Table \ref{table-combin-TE-SE}.
\begin{table*}
\centering
\caption{Summary quality evaluation for four data-sets by combining topic and sentence entropy. R-1: ROUGE-1, R-2: ROUGE-2, R-L: ROUGE-L. TE: Topic Entropy, SE: Sentence Entropy.}
\begin{tabular}{|l|c|c|c|c|}
\hline
\textbf{Selection Criteria}&&\textbf{DUC2001} & \textbf{DUC2002}& \textbf{CNN\&DailyMail}\\
\cline{1-5}
\multirow{2}{*}{High TE \& High SE}&R-1 & 36.94 & 38.65& 29.30\\
\cline{2-5}
&R-2 &11.31& 14.18 &8.2\\
\cline{2-5}
&R-L &33.17& 35.11&26.40\\
\hline
\multirow{2}{*}{High TE \& Low SE}&R-1 &41.04 &44.51&29.33\\
\cline{2-5}
&R-2 &13.44 &17.03&9.16\\
\cline{2-5}
&R-L &36.36 & 39.73 &26.23\\
\hline
\end{tabular}
\label{table-combin-TE-SE}
\end{table*}

\begin{algorithm}[H]
\DontPrintSemicolon
\SetAlgoLined
$Summ\leftarrow\phi $\;
\While{Summ\_Len $<$ Desired\_Summ\_Len}{
Select next highest entropy topic  $\tau$\;
Select the sentence with highest  $\zeta^{\tau}(S)$ not included in $Summ$\;
Add selected sentence to $Summ$ and increment $Summ\_Len$\;
}
\caption{High TE \& High SE}
\label{algo-high-high}
\end{algorithm}

By definition, high entropy semantic units are more informative and hence better candidates for inclusion in the summary. Sentence with high entropy signifies high informativeness since it discusses multiple latent topics. However, presence of important topics in the sentence is also desirable at the same time. Algorithmically, this entails selecting topic with high entropy and identifying high entropy sentence participating in it (Algorithm \ref{algo-high-high}). Iterating the procedure with next highest entropy topic generates the summary of desired length. To rule out the possibility of repetition of sentences, sentence with the next highest entropy is selected in case the highest entropy sentence has already been included in summary.

\begin{algorithm}[H]
\DontPrintSemicolon
\SetAlgoLined
$Summ\leftarrow\phi $\;
\While{Summ\_Len $<$ Desired\_Summ\_Len}{
Select next highest entropy topic  $\tau$\;
Select the sentence with lowest  $\zeta^{\tau}(S)$ not included in $Summ$\;
Add selected sentence to $Summ$ and increment $Summ\_Len$\;
}
\caption{High TE \& Low SE}
\label{algo-high-low}
\end{algorithm}

Even though the aforementioned approach theoretically sounds satisfactory, empirical evaluation reveals the caveat. High entropy sentences selected from high entropy topics do not add much value to the summary because of thin spread of focus over topics leading to low quality summaries. Selecting  the sentence with low entropy (i.e. focused on a topic) from the topic with high entropy is astute since low entropy sentence focuses  the characterizing topic more sharply despite covering   fewer topics (Algorithm \ref{algo-high-low}). 

Table \ref{table-combin-TE-SE} shows that the combination of  \textit{High TE \& High SE} for sentence selection does not perform as well as the combination of \textit{High TE \& Low SE}.  Significant gain in the performance by including low entropy sentences in the summary is somewhat counter intuitive.  Complexity associated with interlacing topic and sentence entropy enjoins clever combination of information contained in topics and sentences.

%%%%%%%%%%%%%%%%%%%%%%%%%%%%%%%%%%%%%%%%%%%
%%%%%%%%%%  SECTION - 5 %%%%%%%%%%%%%%%%%%%
%%%%%%%%%%%%%%%%%%%%%%%%%%%%%%%%%%%%%%%%%%%

\section{Generating Informative Summary}
\label{sec-algorithm}
In this section, we propose an unsupervised algorithm called \textit{E-Summ}, which is an information theoretic method for extractive single document summarization. 
The  method exploits information conveyed by topic and sentence entropies in the latent semantic space of the document to identify candidate sentences. Subsequently, it selects the summary sentences by optimizing the information content while delimiting summary length. \textit{E-Summ} is an explainable, language-, domain- and collection- independent algorithm.

\subsection{Sentence Scoring}
\label{sec-sentence-scoring}
Let $X$ be a random variable with $P(x)$ denoting the probability of occurrence of event $X=x$. \textit{Information} associated with the event $X=x$ is quantified as $I_X(x) = -log_2(P(x))$ and is referred as self-information (\cite{goodfellow2016deep}). This definition asserts that a less likely event is more informative than a highly likely event. Accordingly, a sentence with the highest contribution (probability) in the topic has minimum self-information in that topic. Thus, intuitively a sentence with low self-information is more informative and hence a good representative of the topic. \textit{E-Summ} algorithm uses this principled criterion for selecting \textit{informative} sentences from \textit{important} topics.

Let $\hat\chi_{ij}$ denote the event that sentence $S_{j}$ participates in latent topic $\tau_{i}$, and $\hat{h}_{ij}$ be the probability of this event (Sec. \ref{sec-topic-entropy}). Using the notation for self-information, topic entropy in sentence space (Eq. \ref{eqn-topic-ent-H}) can be rewritten as
\begin{equation}
    \Psi^{S}(\tau_{i}) = \sum\limits_{j=1}^n \hat{h}_{ij} I(\hat\chi_{ij})
\end{equation}

Likewise, let $\chi^{\prime}_{ij}$ denote the event that latent topic $\tau_{i}$ contributes in sentence $S_{j}$ and $h^{\prime}_{ij}$ be the probability of the event  $\chi^{\prime}_{ij}$ (Sec. \ref{sec-sentence-entropy}). Equation \ref{eqn-sent-ent-H} can therefore be rewritten in terms of self-information as follows.
\begin{equation}
    \zeta^{\tau}(S_{j}) = \sum\limits_{i=1}^r h^{\prime}_{ij} I(\chi^{\prime}_{ij})
\end{equation}

Based on the ground rule that a sentence with low self-information is the good representative of a topic, \textit{E-Summ} identifies candidates by choosing sentences with the least self-information from each latent topic and assigns a score as the sum of sentence entropy in topic space ($\zeta^{\tau}(S_{j})$) and topic entropy in sentence space ($\Psi^{S}(\tau_{i})$) as follows.
\begin{equation}
Score(S_{j}) = \zeta^{\tau}(S_{j}) + \Psi^{S}(\tau_{i})
\label{eqn-sentence-score}
\end{equation}

\textit{E-Summ} selects summary sentences from the set of candidate sentences such that the total score of the selected sentences is maximized for the pre-defined summary length. For this purpose, we use the classical \textit{Knapsack} optimization algorithm. Given a set of items, each associated with weight and value, the algorithm selects a subset of items such that the total value of items is maximized for a constant capacity (weight) of the knapsack.

Each candidate sentence identified by \textit{E-Summ} has two associated attributes - sentence length and score. Considering sentence length as item weight, score as item value, and required summary length as the capacity, the \textit{Knapsack} algorithm selects sentences from the set of candidates to maximize the total score for the required summary length. Thus the algorithm maximizes the total information conveyed by chosen summary sentences.

\begin{algorithm}[H]
\DontPrintSemicolon
\SetAlgoLined
% \Input{Term-sentence matrix $A$\, }
% \Output{Document_Summary}
\KwIn{Term-sentence matrix $A$, Number\_of\_topics $r$,  Summ\_Len $\mathcal{L}$}
\KwOut{Document Summary $Summ$ }
$W$, $H$  $\leftarrow$ NMF($A$) \tcp*[l]{Decompose $A$ into $r$ topics}
Calculate entropy $\Psi^S(\tau)$ for each topic \tcp*[l]{Using Eq.\ref{eqn-topic-ent-H}}
Sort topics in decreasing order of entropy\;
$Cand\_Set\leftarrow\phi $, $Summ\leftarrow\phi$ \tcp*[l]{Init candidate set and summary}
%$S\_Len\leftarrow0$ 
%\While{S\_Len $<$ Desired\_Summ\_Len}
\For{$i\gets1$ \KwTo $r$}{
$\tau_{i}\gets$ Select next highest entropy topic  $\tau$\;
$C \gets$Select the sentence with minimum $I(S, \tau_{i})$ \;
$Score(C) = \zeta^{\tau}(C) + \Psi^{S}(\tau_{i})$  \tcp*[l]{Use Eq.\ref{eqn-sentence-score}}
Add $C$ to $Cand\_Set$\;% and increment S\_Len\;
}
$Summ\gets$ Knapsack($Cand\_Set$, $\mathcal{L}$)\;
\If {$length($Summ$)$ $<$ $\mathcal{L}$} 
{
    $Rem\_Sent$ $\gets$ $Summ$ $\setminus$ $Cand\_Set$ \tcp*[l]{candidates not in $Summ$}
    Sort $Rem\_Sent$ in decreasing order of score\; 
    Select top scoring sentences to complete $\mathcal{L}$\;
}
%\EndIf
\caption{E-Summ Algorithm}
\label{proposed-algo}
\end{algorithm}

Algorithm \ref{proposed-algo} presents the pseudo-code for the proposed \textit{E-Summ} algorithm. We pre-process document $D$ by removing stop-words, punctuations and transform it to binary term-sentence matrix $A$. Input parameters to \textit{E-Summ} are matrix $A$, number of latent topics $r$, and desired summary length.

The algorithm  decomposes matrix $A$ into two factor matrices $W$ and $H$  using NMF (step 1). In the interest of stability of NMF factor matrices, we use NNDSVD\footnote{According to our experience with NNDSVD initialization, randomization is not completely eliminated.} initialization proposed by \cite{boutsidis2008svd}. In step 2, the method calculates entropy of each latent topic in sentence space (Sec. \ref{sec-topic-entropy}). Next, \textbf{for} loop in steps 5 to 9, examines latent topics in descending order of their entropy and identifies the best representative sentence (with minimum self-information) for that topic. The algorithm appends the selected sentence to the set of candidates. Subsequently, in step 11, the \textit{Knapsack} optimization algorithm is applied to maximize information content of the summary while delimiting the summary length.

In case the sentences selected by \textit{Knapsack} algorithm do not complete the desired summary length, remaining candidates are considered in decreasing order of their score for inclusion in the summary (steps 12 to 16). However, when desired summary length is specified as number of sentences, application of \textit{Knapsack} algorithm is omitted and top scoring candidates are selected for inclusion in summary.

\textit{E-Summ} algorithm is an unsupervised document summarization method, which does not require external knowledge resources. It is generic, domain- and collection- independent and  does not use language tools at any stage. The algorithm is topic-oriented and hence is highly effective for documents, which have differentiable topics.

The algorithm is efficient with  computational complexity $O(r(n+\mathcal{L}))$,  where $O(rn)$ is the time to extract candidate sentences and $O(n\mathcal{L})$ time is required to execute the knapsack algorithm using dynamic programming. The complexity is linear in the number of latent topics ($r$), number of sentences ($n$), and desired summary length ($\mathcal{L}$) and excludes the  complexity of non-negative matrix factorization (step 1). Though NMF is an  NP-hard problem \citep{NMFcomplexity}, efficient polynomial-time algorithms that use local search heuristic are commonly available in libraries.

\subsection{Explainability of E-Summ}
\label{sec-esumm-explain}
Non-negativity constraint on NMF factor matrices enhances interpretability of latent space and accentuates explainability of  \textit{E-Summ} algorithm. Given the summary generated by the algorithm, it is possible to retrace the selection process and justify inclusion of sentences in  summary. We illustrate the procedure using the document shown in Fig. \ref{fig-example-1-a}, while decomposing it into four topics. 

\begin{figure}[htbp!]
\centering
\caption{Explaining E-Summ summary for Document No. AP881118-0104 from DUC2002 data-set shown in Fig. \ref{fig-example-1-a}}
\label{fig-esumm-explain}
\begin{subfigure}[b]{\textwidth}
\caption{Matrix displaying probability distribution of topics in sentences (top) and self-information of sentences in topics (bottom), $0$ indicates that the sentence does not contribute in the topic, $\Psi^{S}(\tau)$: topic entropy in sentence space. $\ell$: Sentence Length}
\label{fig-esumm-explain-a}
\noindent
%\fbox{
  % \parbox{
\scriptsize{
\begin{equation*}
  \begin{blockarray}{*{12}{c} l}
    \begin{block}{*{12}{c} l}
      S1 & S2 & S3 & S4 & S5 & S6 & S7 & S8 & S9 & S10 & S11 & S12 & \Psi^{S}(\tau)\\
    \end{block}
    \\
    \begin{block}{[*{12}{c} ]l}
    0 & \frac{0.65}{0.63} & \frac{0.02}{5.95} &  \frac{0.05}{4.35} & \frac{0.03}{4.92} &\frac{0.01}{6.80} & \frac{0.02}{5.49} & 0 &\frac{0.09}{3.55} &\frac{0.14}{2.87}&0&0& \tau_1  (1.76)\\
    \\
    \frac{0.50}{0.10}	& 0 &0 &\frac{0.04}{4.60} &\frac{0.02}{5.64} &\frac{0.33}{1.60} &	\frac{0.08}{3.74} &\frac{0.03}{5.25} &	0 &\frac{0.01}{7.03}&0&0& \tau_2  (1.80)\\
    \\
    0 &	0 &	0 & \frac{0.12}{3.06} &0 &\frac{0.04}{4.55} &0 & 0 &0 &\frac{0.00}{7.98}&0&\frac{0.83}{0.26}& \tau_3  (0.81)\\
    \\
    0 & 0&\frac{0.29}{1.79} & 0 & \frac{0.15}{2.77} & 0 & \frac{0.10}{3.32} & \frac{0.01}{6.60} & \frac{0.01}{7.57} &0&\frac{0.45}{1.16}&0& \tau_4  (1.88)\\ \\
    0&	0	&0.32 &	1.52	&1.10&	0.59&	1.40&	0.88&	0.30	&0.39	&0	&0&\zeta^{\tau}(S)\\
    29&28&15&	16&	12&	24&	14&	7&	9&	10&	22&	24 &\ell\\
    \end{block}
    \end{blockarray}
    \end{equation*}
}
\end{subfigure}

\begin{subfigure}[b]{\textwidth}
\caption{E-Summ Summary}
\label{fig-esumm-explain-b}
\noindent\fbox{%
    \parbox{\textwidth}{%
\scriptsize{\setstretch{1.0}{
\textbf{S11:} Since Sept. 19, when he was confined to intensive care after vomiting blood, Hirohito has received 42 pints of blood in transfusions.

\textbf{S1:} Emperor Hirohito has lost nearly 40 pounds in the two months since he fell ill and now weighs only 66 pounds, a palace official was quoted as saying Friday.

\textbf{S2:} The 87-year-old emperor's condition weakened Friday, but doctors speculated that a scab had stopped internal bleeding in his upper intestinal area, said Imperial Household Agency spokesman Kenji Maeda.

\textbf{S12:} Palace officials have not given details on Hirohito's illness, and they would not comment on Japanese news reports that the emperor has pancreatic cancer.
}}
}
}
\end{subfigure}

\begin{subfigure}[b]{\textwidth}
\centering
\caption{Explaining selection of candidate sentences. \textbf{Candidates}: Candidate sentences identified by E-Summ, \textbf{$\ell$}: Sentence Length, Topics  listed in decreasing order of entropy, \textbf{I(S,$\tau$)}: Minimum self-information of sentence $S$ in topic $\tau$,\textbf{ Score(S)}: Sentence score (Eqn. \ref{eqn-sentence-score})}
\label{fig-esumm-explain-c}
\scriptsize{
\begin{tabular}{|p{5em}|p{5em}|p{5em}|p{5em}|p{8em}|}
\hline
\textbf{Candidates}&\textbf{$\ell$}&\textbf{Topics} & \textbf{$I(S,\tau)$}&\textbf{Score (S)}\\
\hline
$S11$&22&$\tau_4$& 1.16&1.88 (0.0 + 1.88)\\
\hline
$S1$&29&$\tau_2$&1.00&1.80 (0.0 + 1.80)\\
\hline
$S2$&28&$\tau_1$&0.63&1.76 (0.0 + 1.76)\\
\hline
$S12$&24&$\tau_3$&0.26&0.81 (0.0 + 0.81)\\
\hline
\end{tabular}
}
\end{subfigure}

\begin{subfigure}[b]{\textwidth}
\centering
\caption{Justifying selection of summary sentences}
\label{fig-esumm-explain-d}
\scriptsize{
\begin{tabular}{|l|l|}
\hline
Knapsack sentences&S1, S2, S11\\
\hline
Maximum score &5.44\\
\hline
Length of Knapsack sentences&79 words\\
\hline
Sentence added to complete summary & S12\\
\hline
\end{tabular}
}
\end{subfigure}
\end{figure}

\begin{Ex} We elucidate the transparency of proposed \textit{E-Summ} method using the document of Example \ref{exam-doc}.  Fig. \ref{fig-esumm-explain-a} presents the matrix specifying probability of participation of topics in sentences and self-information  of sentences in the document. The top component of element ($i, j$) denotes the probability of participation of topic $\tau_i$ in sentence $S_j$($\hat{h}_{ij}$) and bottom component denotes the self-information of sentence $S_j$ in topic $\tau_i$ ($I(S_{j}, \tau_{i})$). The element with value $0$ indicates that the sentence does not contribute in the topic. The two rows below the matrix show sentences' entropies and lengths respectively.

Fig. \ref{fig-esumm-explain-b} shows  summary of the document (Fig. \ref{fig-example-1-a}), with sentences listed in order in which they are selected by \textit{E-Summ}. Fig. \ref{fig-esumm-explain-c} shows the order of selection of candidate sentences. \textit{E-Summ} algorithm starts with the topic of highest entropy, $\tau{4}$ and selects its best representative sentence $S11$. Continuing the process with remaining topics (in order), the algorithm adds sentences $S11$, $S1$, $S2$ and $S12$ to the candidate set. Applying \textit{Knapsack} algorithm on candidate sentences extracts $S1$, $S2$ and $S11$ for inclusion in summary by maximizing the total sentence score for summary limit of $100$ words (Fig. \ref{fig-example-1-d}). However, total length of sentences  $S1$, $S2$ and $S11$ is $79$ words. Since desired summary length is not attained, \textit{E-Summ} algorithm considers remaining candidate sentence (only $S12$ in this case) to complete the summary length.
\blackslug
\end{Ex}
%%%%%%%%%%%%%%%%%%%%%%%%%%%%%%%%%%%%%%%%%%%
%%%%%%%%%%  SECTION - 6 %%%%%%%%%%%%%%%%%%%
%%%%%%%%%%%%%%%%%%%%%%%%%%%%%%%%%%%%%%%%%%%

\section{Experimental Design}
\label{sec-experimental-design}
In this section, we describe the experimental setup required for the performance evaluation of the proposed \textit{E-Summ} algorithm. The algorithm  is implemented in Python $3.7$ using Natural Language Toolkit (NLTK), \textit{textmining} package and \textit{Scikit-learn} toolkit. All experiments are carried out on an Intel(R) Core(TM) i5-8265U CPU running Windows 10 OS with 8GB RAM.

\subsection{Data-sets}
\label{sec-datasets}
We use four well known data-sets DUC2001\footnotemark, DUC2002\footnotemark[\value{footnote}]\footnotetext{\url{http://duc.nist.gov}}, CNN\footnotemark  and DailyMail\footnotemark[\value{footnote}]\footnotetext{CNN and DailyMail corpora contain news articles and  were originally constructed by \cite{hermann2015teaching} for the task of passage-based question answering, and later re-purposed for the task of document summarization.} to evaluate  quality of  summaries generated by \textit{E-Summ} algorithm. DUC2001 and DUC2002 data-sets comprise of $308$ and $533$ documents respectively. Each document in these data-sets has $1-3$ abstractive reference summaries of approximately $100$ words. CNN and DailyMail data-sets are divided into training, validation and test sets with $92,579\, (90,266 / 1,220 / 1,093)$ and $219,506\,(196,961 / 12,148 / 10,397)$ documents respectively. Each document in these two data-sets is accompanied by one reference summary consisting of story highlights. Following previous research work (\cite{narayan2018ranking, al2018hierarchical, zhou2018neural}), we extract three sentences for a CNN document summary and four sentences for a DailyMail document summary for comparative performance evaluation.  Interestingly, all four data-sets consist of news articles. In order to establish the claim of domain independence of \textit{E-Summ}, we evaluate the performance of \textit{E-Summ} algorithm on well known CL-SciSumm 2016 data-set \citep{clscisummtask-2016} for scientific document summarization.

\subsection{Number of Latent Topics}
\label{sec-topics}
Since \textit{E-Summ} selects  representative sentences from  important topics in the latent semantic space, decomposing the document into optimal number of topics is critical to  the quality of summary produced by the algorithm. In case the number of sentences desired in summary is known, setting dimensionality of the latent semantic space is straightforward. For example, number of sentences desired for CNN/DailyMail data-sets summaries is three/four, respectively. Accordingly, we set the number of latent topics ($r$) to $3$ for CNN data-set and $4$ for DailyMail data-set, based on the assumption that each sentence briefs one topic. In other scenarios, finding the optimal number of latent topics is a tricky task because it is an unknown and complex function of writing style adopted by the author,  desired length of the summary,  and the background knowledge of the reader.

The number of topics in a document corresponds to the number of core concepts (topics) or ideas described in the document. Latent Dirichlet Allocation (LDA), which is a well-known generative probabilistic method for topic modeling, is unsuitable because it requires number of latent topics as an input parameter \citep{LDA_blei2003latent}. Therefore, we propose to find the number of latent topics by employing a community detection algorithm as described below.

%for topic modeling, which represents a document as random mixture over latent topics, and characterizes each topic by a distribution over words (\cite{LDA_blei2003latent}). However,  the method suffers from the bane of parameter selection \textbf{including the number of topics} and does not justify the use of Dirichlet prior in the model formulation (\cite{gerlach2018network}). 

A concept (topic) in the document is communicated by a group of semantically related and frequently co-occurring terms. These groups manifest as communities in the co-occurrence  graph of the document, wherefore  identifying  communities  in this network  can potentially reveal the optimal number of topics in the document. The idea of segmenting the document into latent topics using word co-occurrence network representation of the text has been endorsed in recent works on topic modeling \citep{kido2016topic, dang2018commodeler, gerlach2018network}.

Consider the term-sentence matrix $A$ for  document $D$. Then,  $ Q = AA^T$ is the  term-term matrix for $D$, where element $q_{ij}$ denotes the number of the times term $t_{i}$ occurs with term $t_{j}$ in the document. Considering matrix $Q$ as the weighted adjacency matrix representation of the co-occurrence graph of $D$, application of community detection algorithm reveals the groups of terms related in the latent semantic space. Each   group (community) maps to a topic discussed in the document. We use  Louvain algorithm \citep{blondel2008fast} for detecting communities, a simple and fast algorithm to detect communities, and is based on the heuristic of optimizing modularity.

\begin{figure}[h!]
\begin{minipage}{0.45\textwidth}
\includegraphics[width=0.95\textwidth]{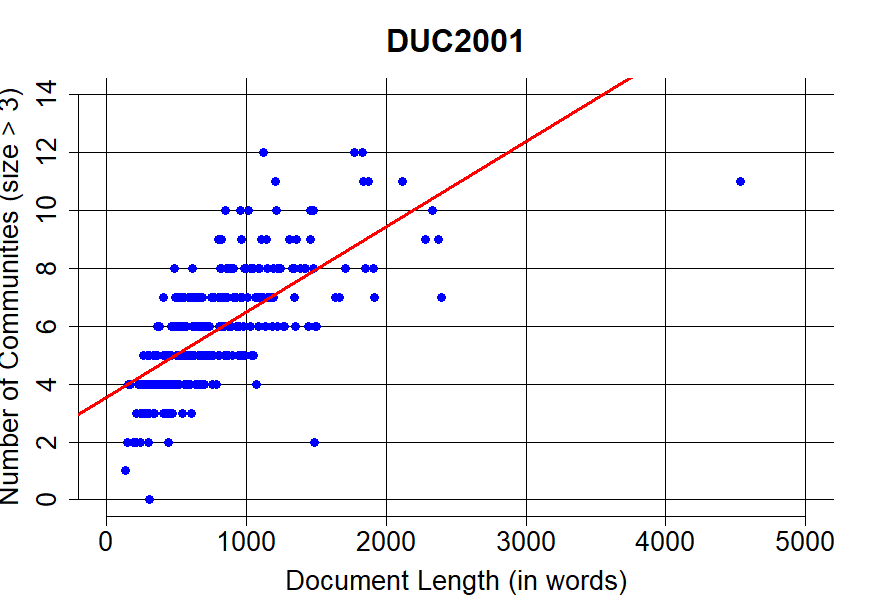}
\end{minipage}
\begin{minipage}{0.45\textwidth}
\includegraphics[width=0.95\textwidth]{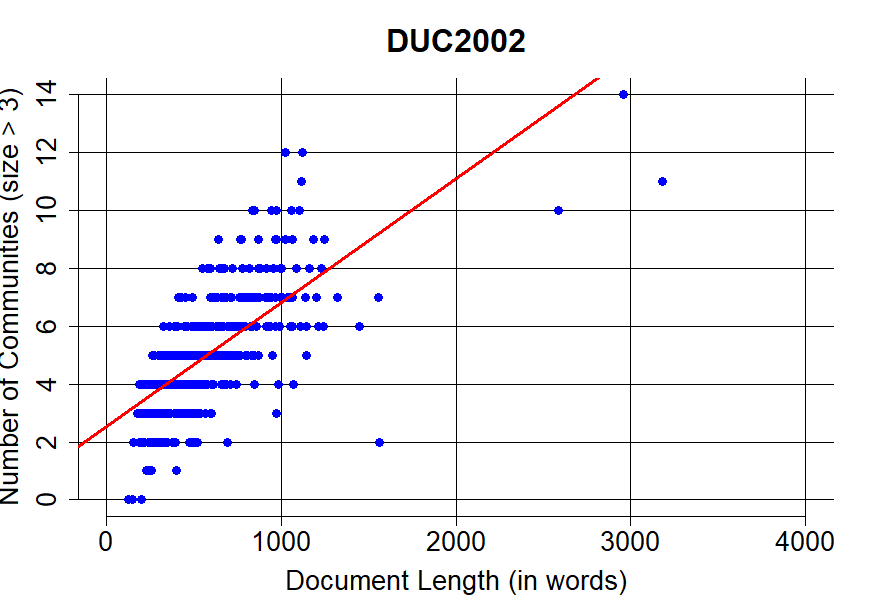}
\end{minipage}
\caption{Correlation between number of communities and document length for DUC2001 and DUC2002 data-sets}
\label{fig-commsize-doclen-correl}
\end{figure}

The number of communities and their sizes are specific to the content and writing style of the document. Longer documents often describe more ideas or topics and therefore reveal more communities compared to shorter documents. We found a healthy correlation of 0.70 between the document length and the number of communities in DUC2001 and DUC2002 (Fig. \ref{fig-commsize-doclen-correl}). The deviant behaviour of  some documents in both data-sets is due to atypical content and writing style.

The size of the community is suggestive of the extent of coverage of the topic in the document. Larger communities possibly span multiple sentences, forging strong candidature for inclusion in summary.  On the other hand, smaller communities denote  author ideas that are feeble and limited to fewer sentences.  We consider a community of size less than four to be an artifact indicating spurious  co-occurrence and do not recognize it as a valid topic.

The above stated  heuristic sometimes results into  inadequate number of communities  to complete the summary (fewer than three\footnote{We envisage that this threshold may vary with the desired summary length.}). In such a situation, the number of latent topics is computed by taking into account the required summary length and the average sentence length as follows.
\begin{equation}
    r =\frac{\text{Required Summary Length}}{\text{Average Sentence Length}}
\label{eqn-latent-topics}
\end{equation}

\noindent
To summarize the above discussion, we recommend the   following methods to determine the value of number of latent topics, $r$.
\begin{enumerate}[label=(\roman*)]
    \item When the summary length is given as number of sentences, $r$ is set to  desired summary length.
    \item When the summary length is specified in number of words, $r$ is computed using community detection method described above.
    \item When the community detection method results in an inadequate number of communities, $r$ is computed using Eq. \ref{eqn-latent-topics}.
\end{enumerate}

\subsection{Competing Methods}
\label{sec-competing-methods}
To the best of authors' knowledge, no published work for single document extractive summarization evaluates performance on all four data-sets. For comparison with the state-of-the-art, we group algorithms based on the data-sets on which they are evaluated. Accordingly, for DUC2001 data-set, COSUM (\cite{alguliyev2019cosum}) and method by \cite{saini2019extractive} are used as competing methods and for DUC2002, CoRank+ (\cite{fang2017wordcorank}) and method by \cite{saini2019extractive} are used. Interestingly, no unsupervised method is evaluated on CNN and DailyMail data-sets. We, therefore, compare the performance of \textit{E-Summ} against supervised methods for these two data-sets. We use our earlier  work  on NMF based document summarization as baseline performance (\cite{khurana2019extractive}). \textit{E-Summ} performance on scientific and generic articles is compared with recent algorithms detailed in Section \ref{sec-scientific-doc-summarization}.

\subsection{Evaluation Metrics}
\label{sec-evaluation-metrics}
ROUGE  is the de-facto metric for qualitative assessment of automatic summarization algorithms, and ROUGE toolkit is the most commonly used software package for this purpose \citep{lin2004rouge}. ROUGE performs \textit{content-based} evaluation by matching uni-grams, bi-grams and n-grams between system and reference (human-produced) summaries.  It  generates three metrics viz. recall, precision, and F-measure while evaluating system summary against reference summary. 

Following previous studies, we compute  recall metric for evaluating DUC2001, DUC2002 documents summaries and F-measure for evaluation of CNN, DailyMail documents summaries. All reported ROUGE scores are macro-averaged over the data-set. For each ROUGE metric, we use three variations viz. ROUGE-1 (R-1), ROUGE-2 (R-2) and ROUGE-L (R-L) for performance evaluation. R-1 finds overlapping uni-grams, R-2 identifies bi-grams, and R-L locates longest common n-gram between system and reference summaries. 

Difference between the vocabularies of system and reference summaries underestimates the ROUGE score. Further, ROUGE does not work when reference summaries are not available. During the last two years, there has been a spurt in research related to metrics for summary quality \citep{peyrard2019studying, bhandari2020metrics, huang2020achieved, vasilyev2020human, fabbri2020summeval, bhandari2020re-evaluation}. Most of  these works have argued against the ROUGE metric because it fails to robustly match paraphrases resulting in misleading scores, which do not correlate well with human judgements  \citep{zhang2019bertscore, huang2020achieved}. \citet{MATCHSUM_zhong2020extractive} argue that high semantically similarity with the source document is highly desirable for a  good summary.

\subsubsection{Semantic Similarity}
\label{sec-semantic-similarity}
The current churn in the summary quality evaluation metrics prompts us to employ semantic similarity as an additional measure. Typically, reference  summaries  are produced by humans and hence are abstractive\footnote{All four data-sets commonly used for evaluation of extractive summarization systems sport abstractive summaries as reference.}.  \cite{lin2002manual} observe low inter-human agreement of approximately $40\%$ in single document summarization task and $29\%$ in multi-document summarization task for DUC2001 data-set. The authors advocate evaluation against more reference summaries for higher confidence in the quality of extractive summaries. 

Consequent to the above observations,  we supplement our investigation by  evaluating system summaries using semantics based automatic evaluation measure proposed by \cite{steinberger2012evaluation}. This method evaluates system summaries w.r.t original document instead of reference summaries to mitigate the challenges related to - (i) few reference summaries, and (ii) difference in the vocabulary of system and reference summaries. 

\cite{steinberger2012evaluation} propose Latent Semantic Analysis (LSA) based \textit{content} evaluation measure to evaluate a summary w.r.t the original document. LSA is a Singular Value Decomposition (SVD) based technique, which reveals latent semantic space of the document. We briefly describe the \textit{content-based} LSA measure for summary evaluation below.

Given the binary term-sentence matrix $A_{m \times n}$, application of SVD yields $A = U \sum V^{T}$, where $U =[u_{ij}] $ is $m \times r$ term-topic matrix, $\sum =[\sigma_{ij}]$ is $r \times r$ diagonal matrix  and $V^{T}$ is a $r \times n$ topic-sentence matrix (here, $r = \min\{m,n\}$). The diagonal elements in $\sum$ denote the importance of topics in descending order. For computing the importance of terms in latent space, the method computes matrix $B$ as follows.

\begin{equation*}
B_{m,r} = 
\begin{pmatrix}
u_{11}\sigma_{1}^{2} & u_{12}\sigma_{2}^{2} & \cdots & u_{1r}\sigma_{r}^{2} \\
u_{21}\sigma_{1}^{2} & u_{22}\sigma_{2}^{2} & \cdots & u_{2r}\sigma_{r}^{2} \\
\vdots  & \vdots  & \ddots & \vdots  \\
u_{m1}\sigma_{1}^{2} & u_{m2}\sigma_{2}^{2} & \cdots & u_{mr}\sigma_{r}^{2} 
\end{pmatrix}
\end{equation*}

Each element $b_{ij}$ in matrix $B$ quantifies the contribution of term $t_{i}$ weighted by the importance of the latent topic $\tau_{j}$. The overall importance of $t_i$ is computed as $\mid b_{i} \mid = \sqrt{b_{i1}^{2} + b_{i2}^{2} + \ldots + b_{ir}^{2}}$. The term vector for complete document is given as 
$\begin{bmatrix}
\mid b_{1} \mid, \mid b_{2} \mid, \ldots, \mid b_{m} \mid
\end{bmatrix}^T$.

The term vector is computed for system summary and the original document. The cosine of angle between two resulting vectors represents semantic similarity between system summary and the original document. 

\begin{figure}[h]
\centering
\includegraphics[width=\textwidth]{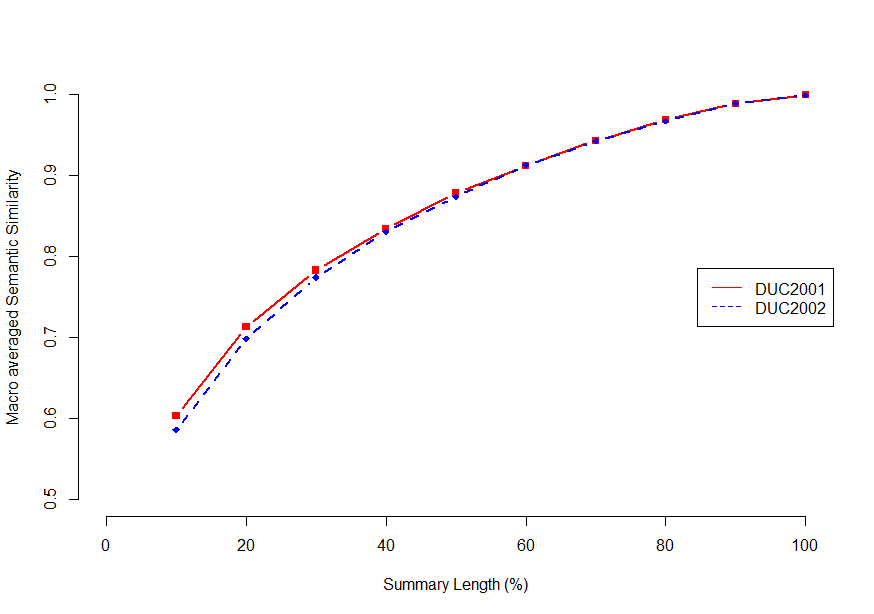}
\caption{Macro-averaged semantic similarity of E-Summ summaries w.r.t corresponding original documents for summary length as percentage of document size (DUC2001 and DUC2002)}
\label{fig-lineplot-duc}
\end{figure}

We  examine the efficacy of semantic similarity as a measure  of automatic evaluation of summary quality. Using DUC2001 and DUC2002 data-sets, we generate \textit{E-Summ} summaries of varying lengths (10\%, 20\%, \ldots,100\% of original document) for each document. Next, we compute the semantic similarity of \textit{E-Summ} summaries w.r.t original documents and plot the macro-averaged values in  Fig. \ref{fig-lineplot-duc}. It is clearly seen that  increasing summary length improves semantic similarity, approaching $1$  for 100\% summary length. This experiment validates the efficacy of semantic similarity as a measure for evaluating  quality of the summary.

%%%%%%%%%%%%%%%%%%%%%%%%%%%%%%%%%%%%%%%%%%%
%%%%%%%%%%  SECTION - 7 %%%%%%%%%%%%%%%%%%%
%%%%%%%%%%%%%%%%%%%%%%%%%%%%%%%%%%%%%%%%%%%

\section{Performance Evaluation}
\label{sec-experimental-results}
We report ROUGE scores of summaries, macro-averaged over all documents in the collection, along with respective standard deviations. We present dataset-wise comparative evaluation of \textit{E-Summ} algorithm in Sections  \ref{sec-perf-duc2001} to \ref{sec-perf-cnn-dm},  followed by  quantitative assessment of  \textit{E-Summ} summaries using  semantic similarity  in Section  \ref{sec-sem-sim-results}. We  substantiate the claim of language independence in Section  \ref{sec-language-agnostic}.

\subsection{Performance on DUC2001 data-set}
\label{sec-perf-duc2001}
Table \ref{table-duc2001-recall} shows results of comparative evaluation of \textit{E-Summ} algorithm on DUC2001 data-set. To the best of our knowledge, no deep neural method for  document summarization  has been evaluated on DUC2001 data-set. This  limits comparison of  the performance evaluation of \textit{E-Summ} algorithm with only unsupervised methods. 

\begin{table}[!h]
\scriptsize
\centering
\caption{Performance comparison of \textit{E-Summ} algorithm for DUC2001 data-set based on ROUGE recall scores, reported along with  standard deviation. Best performance is in boldface.}
\begin{tabular}{|l|l|c|c|c|}
\hline
&&\textbf{ROUGE-1} & \textbf{ROUGE-2} & \textbf{ROUGE-L}\\
\hline
\multirow{2}{4em}{Baseline Methods} 
&NMF-TR (2019)&44.7 $\pm$ 0.1 & 15.9 $\pm$ 0.1&39.3 $\pm$ 0.1\\
&NMF-TP (2019)& 43.7 $\pm$ 0.1&15.6 $\pm$ 0.1 &38.5 $\pm$ 0.1\\
\hline
{Proposed Method}
&E-Summ& 45.6 $\pm$ 0.1 & 15.71 $\pm$ 0.1& \textbf{40.22} $\pm$ 0.1\\
\hline
\multirow{2}{6em}{Unsupervised Methods}&TextRank (2004) & 43.71&16.63&38.77\\
&LexRank (2004)&42.15&15.31&37.54\\
&COSUM (2019) &47.3 & 20.1 &-\\
&\cite{saini2019extractive} &\textbf{50.24}& \textbf{29.24} &-\\
\hline
\end{tabular}
\label{table-duc2001-recall}
\end{table}

Table \ref{table-duc2001-recall} shows that the quality of summaries generated by \textit{E-Summ} algorithm is better than those generated by the baseline methods, TextRank \citep{mihalcea2004textrank} and LexRank \citep{erkan2004lexrank} methods except for slightly degraded R-2 performance. TextRank and LexRank are two classical algorithms\footnote{\label{footnote-miso}We use the implementation available at \url{https://github.com/miso-belica/sumy}.} for extractive document summarization and present strong baselines. Next, we choose two recent unsupervised methods, which co-incidentally formulate document summarization as an optimization problem. Both methods are collection- and domain- independent like \textit{E-Summ}, but extensively use language tools.

The first method, COSUM, formulates document summarization as clustering based optimization problem and employs an adaptive differential evolution algorithm to solve it (\cite{alguliyev2019cosum}). The method identifies topics in the document and clusters sentences using k-means algorithm. Subsequently, it selects sentences from clusters using an objective function that maximizes coverage and diversity in summary. The second method, proposed by \cite{saini2019extractive}, employs multi-objective binary differential evolution based optimization strategy to generate document summary. The method simultaneously optimizes multiple summary attributes such as similarity of sentences with document title, position of summary sentences in the document, sentence length, cohesion, and coverage of summary. Both methods use variation of genetic approach to find a global optimum solution.

Table \ref{table-duc2001-recall} reveals that recent competing methods (\cite{saini2019extractive, alguliyev2019cosum}) have higher R-1 and R-2 scores than \textit{E-Summ} algorithm but do not report R-L score. Further, both methods require high computational effort to obtain the optimal solution. The average running time per document reported by \cite{saini2019extractive} for this data-set is $32$ seconds  (Page 20 of the reference), which \textit{excludes} the time required for computation of sentence similarity. Running time for COSUM method is not reported by the authors. We noted that running time per document for \textit{E-Summ} as $0.128$\footnote{Reported configuration of the machine used by \cite{saini2019extractive} is superior to the one used in our experiments.} second for this data-set, averaged over five runs. This \textit{includes} the time required to compute number of latent topics using community detection method and extraction of summary sentences using Knapsack algorithm.

\subsection{Performance on DUC2002 data-set}
\label{sec-perf-duc2002}
Performance evaluation of \textit{E-Summ} algorithm for DUC2002 data-set is presented in Table \ref{table-duc2002-recall}. Though the proposed algorithm performs comparably to the baseline methods, it has mixed performance compared to unsupervised and deep neural methods.

\begin{table}[!h]
\scriptsize
\centering
\caption{Performance comparison of the proposed method for DUC2002 data-set based on ROUGE recall scores reported along with standard deviation. Best performance is in boldface.}
\begin{tabular}{|l|l|c|c|c|c|}
\hline
&&\textbf{ROUGE-1} & \textbf{ROUGE-2}& \textbf{ROUGE-L}\\
\hline
\multirow{2}{6em}{Baseline Methods}
&NMF-TR (2019)& 49.0 $\pm$ 0.1&21.5 $\pm$ 0.1 &44.1 $\pm$ 0.1\\
& NMF-TP (2019) &47.6 $\pm$ 0.1 & 19.7 $\pm$ 0.1&42.4 $\pm$ 0.1\\
\hline
{Proposed Method} 
&E-Summ& 50.39 $\pm$ 0.1 & 21.16 $\pm$ 0.1& 45.19 $\pm$ 0.1\\
\hline
\multirow{2}{6em}{Unsupervised Methods}& TextRank (2004) &48.33&22.54&43.75\\
&LexRank (2004) &45.84&20.18&41.38\\
&CoRank+ (2017)& \textbf{52.6} &25.8 & 45.1\\
&COSUM (2019)&49.08 & 23.09&-\\
&\cite{saini2019extractive}& 51.66 & \textbf{28.85}&-\\
\hline
\multirow{2}{6em}{Deep Neural Methods}
&NN-SE (2016)& 47.4 & 23.0 & 43.5\\
&SummaRuNNer (2017)& 46.6 $\pm$0.8 & 23.1 $\pm$0.9 & 43.03 $\pm$0.8\\
&HSSAS (2018)& 52.1& 24.5 &\textbf{48.8}\\
&DQN (2018)& 46.4 & 22.7 & 42.9\\
\hline
\end{tabular}
\label{table-duc2002-recall}
\end{table}

We observe that \textit{E-Summ} algorithm performs better than both TextRank and LexRank methods except for R-2 performance of TextRank. Algorithm CoRank+, which is language independent like the \textit{E-Summ} and uses a graph based approach, performs better than the proposed algorithm. CoRank+  augments sentence-sentence and word-sentence relationship in the document, which proves to be an advantageous strategy (\cite{fang2017wordcorank}).  R-1 score of \textit{E-Summ} is better than that of COSUM (\cite{alguliyev2019cosum}) method, while R-2 score is lower.  Algorithm proposed by \cite{saini2019extractive} has better R-1 and R-2 performance compared to \textit{E-Summ} algorithm. Though both COSUM and method by \cite{saini2019extractive} perform better than the proposed method, their edge over \textit{E-Summ} is unclear in absence of evaluation report on other data-sets and R-L metric. 

CoRank+ reports average running time per document for DUC2002 data-set to be  $30$ seconds  (Sec. 4.3 of \cite{fang2017wordcorank}). Average running time per document reported by  \cite{saini2019extractive} for this data-set is $20$ seconds  (Page 20 of the reference), excluding time to compute sentence similarity. \textit{E-Summ} on an average summarizes a DUC2002 data-set document in $0.105$ second including the time to compute number of latent topics and extracting summary sentences using Knapsack algorithm. The reported execution time is averaged over five runs.

It is also evident from Table \ref{table-duc2002-recall} that between HSSAS (\cite{al2018hierarchical}) and \textit{E-Summ}, former is the clear winner. However, compared to the other three methods (\cite{cheng2016neural,nallapati2017summarunner, yao2018deepreinforcment}), \textit{E-Summ} has mixed performance.  Considering that all four deep neural methods are trained and tuned on CNN/DailyMail data-set, their cross-collection performance is remarkable. 

\subsection{Performance on CNN and DailyMail data-sets}
\label{sec-perf-cnn-dm}
Table \ref{table-cnnDm-fmeasure} presents the results of comparative evaluation of \textit{E-Summ} algorithm on combined CNN and DailyMail data-sets. Performance of \textit{E-Summ} is better than baseline NMF-TP method but lower than that of NMF-TR method. The reason is that  \textit{E-Summ} being a topic-oriented method, picks sentences from the most informative topics, while CNN and DailyMail documents being news articles, do not have clearly demarcated topics. Consequently, \textit{E-Summ} suffers the disadvantage compared to NMF-TR, which being  term-oriented (Sec. 7 of \cite{khurana2019extractive}), is able to capture better summary sentences.

\begin{table}[!h]
\scriptsize
\centering
\caption{Performance comparison of the proposed method for combined CNN and DailyMail data-sets based on ROUGE F-measure scores reported along with standard deviation. Best performance is in boldface.}
\begin{tabular}{|l|l|c|c|c|c|c|c|c|}
\hline
&&\textbf{ROUGE-1} &\textbf{ ROUGE-2} & \textbf{ROUGE-L}\\
\hline
\multirow{2}{6em}{Baseline Methods}& NMF-TR (2019)& 34.2 $\pm$ 0.1 &13.2 $\pm$ 0.1 &31.0 $\pm$ 0.1\\
& NMF-TP (2019)& 30.4 $\pm$ 0.1& 10.9 $\pm$ 0.1&27.4 $\pm$ 0.1\\
\hline
{Proposed Method} &E-Summ& 30.97 $\pm$ 0.1& 10.94 $\pm$ 0.1 &27.78 $\pm$ 0.1\\
\hline
\multirow{2}{6em}{Unsupervised Methods}& TextRank (2004) &31.88&11.80&28.74\\
&LexRank (2004) &33.32&	11.71&30.00\\
\hline
\multirow{4}{6em}{Deep Neural Methods}
&NEUSUM (2018)&41.59& 19.01& 37.98\\
&REFRESH (2018)&40.0 &18.2& 36.6\\
&HSSAS (2018)& 42.3 & 17.8 &37.6\\
&DQN (2018)& 39.4 &16.1 &35.6\\
&BANDITSUM (2018)&41.5& 18.7 &37.6\\
&SemSim (2020)& 44.72& 21.46&41.53\\
&GSum (2021)&\textbf{45.94} &\textbf{22.32}& \textbf{42.48}\\
\hline
\end{tabular}
\label{table-cnnDm-fmeasure}
\end{table}

In absence of evaluation of recent unsupervised extractive single document summarization methods on CNN and DailyMail data-sets, we compare the performance of \textit{E-Summ} with TextRank and LexRank algorithms. We also choose recent Neural summarization algorithms in this category - NEUSUM \citep{zhou2018neural}, REFRESH \citep{narayan2018ranking}, HSSAS \citep{al2018hierarchical}, DQN \citep{yao2018deepreinforcment}, BANDITSUM \citep{dong2018banditsum}, SemSim \citep{SemSimyoon2020learning} and GSum \citep{dou2021gsum} for comparison.

Table \ref{table-cnnDm-fmeasure} reveals that there is a marginal quality  gap between \textit{E-Summ} and TextRank and LexRank summaries. As expected, all deep neural methods demonstrate significantly better performance than \textit{E-Summ}. We recognize that advances in deep neural methods have been a powerful driver of NLP research in recent years and have particularly benefited automatic text summarization on the benchmark data-sets. We discuss the insights gained from this experiment in Sec. \ref{sec-sem-sim-results} and Sec. \ref{sec-discussion}. We report per document running time of \textit{E-Summ} as $0.087$ second, averaged over five runs. Recall that the required summary length for these data-sets is given as number of sentences. Consequently, community detection and Knapsack algorithms  are omitted  and \textit{E-Summ} executes  expeditiously.

\subsection{Semantic Similarity with Original Document}
\label{sec-sem-sim-results}

In this section, we assess the quality of \textit{E-Summ} summaries using content-based semantic similarity method described in Sec. \ref{sec-semantic-similarity}.

\begin{figure}[h]
\centering
\begin{subfigure}{0.48\textwidth}
\centering
\includegraphics[width=\textwidth]{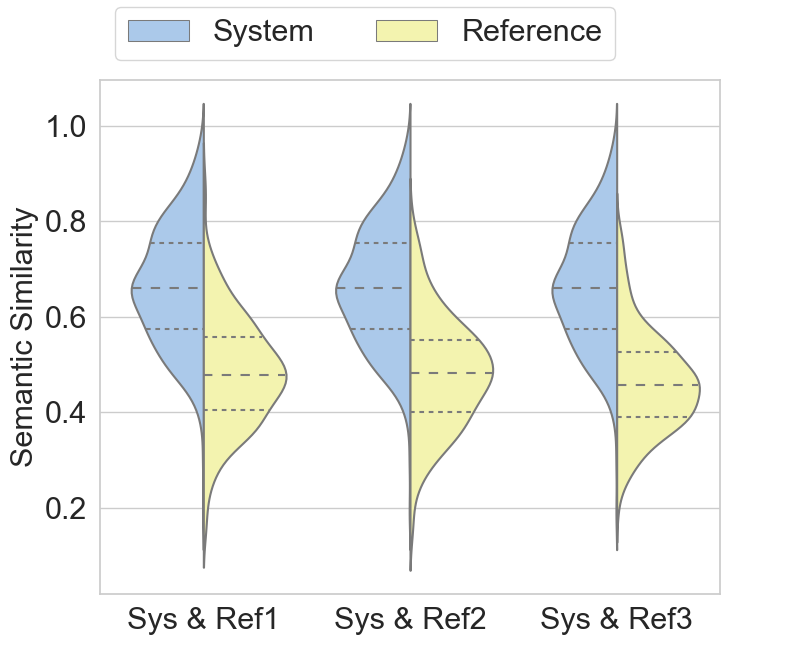}
\caption{DUC2001}
\label{fig-violinplot-duc2001}
\end{subfigure}
\begin{subfigure}{0.48\textwidth}
\centering
\includegraphics[width=\textwidth]{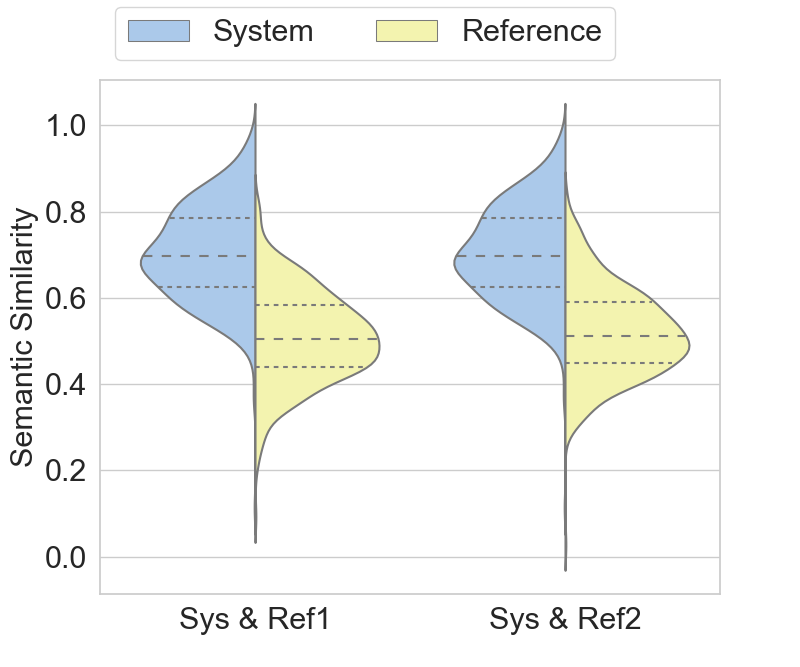}
\caption{DUC2002}
\label{fig-violinplot-duc2002}
\end{subfigure}
\begin{subfigure}{0.48\textwidth}
\centering
\includegraphics[width=\textwidth]{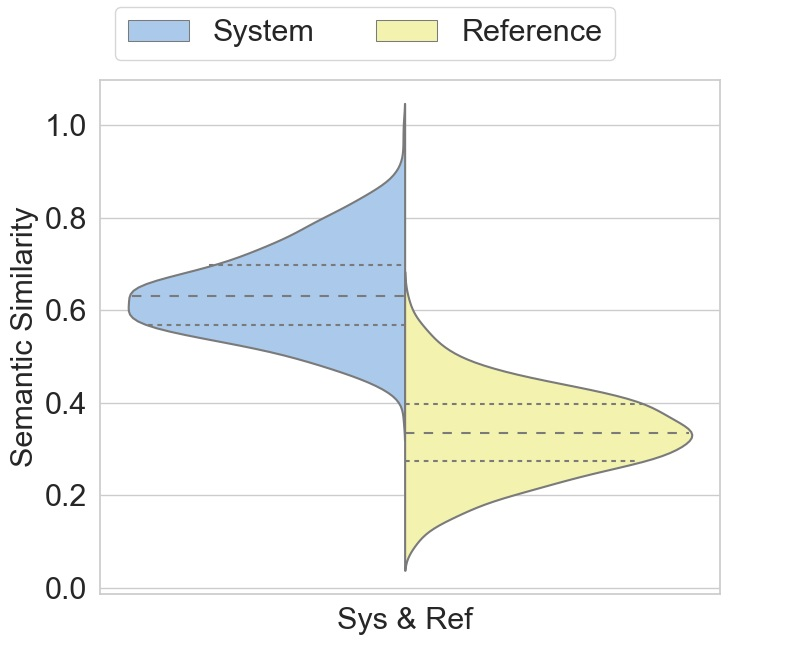}
\caption{CNN}
\label{fig-violinplot-cnn}
\end{subfigure}
\begin{subfigure}{0.48\textwidth}
\centering
\includegraphics[width=\textwidth]{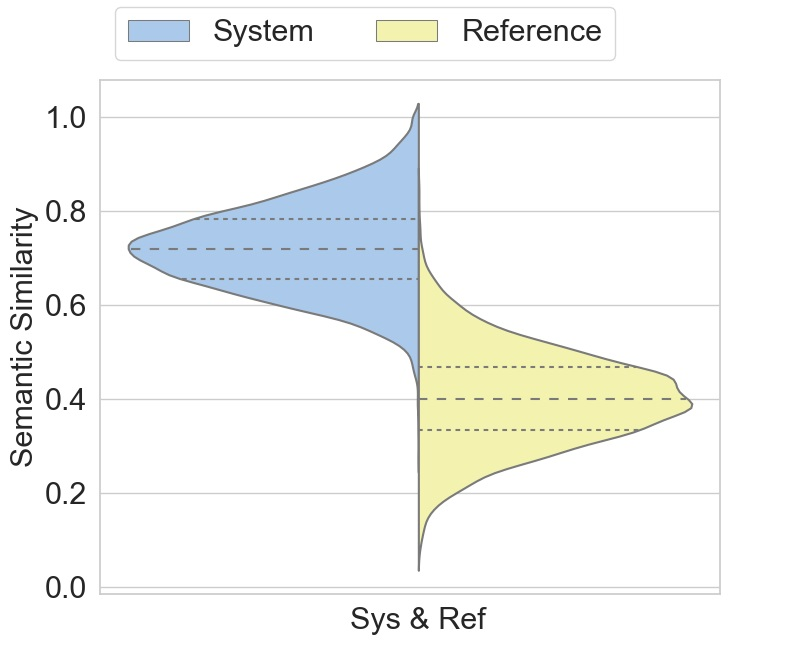}
\caption{DailyMail}
\label{fig-violinplot-DailyMail}
\end{subfigure}
\caption{Comparison of semantic similarity of E-Summ and  reference summaries w.r.t original document for four data-sets. Doc: Document, Sys: E-Summ Summary, Ref: Reference Summary}
\label{fig-violinplot}
\end{figure}

Semantic similarity of system (\textit{E-Summ})  and reference summaries is computed with respect to the original text for each document in all four data-sets. Plots in Figs. \ref{fig-violinplot-duc2001} to \ref{fig-violinplot-DailyMail}  show comparative distributions of similarity scores for system and reference summaries. The plots clearly  reveal that for all data-sets the minimum, maximum, average, first and third quartile scores for reference summaries are lower than those for  system summaries. It is therefore reasonable to conclude that \textit{E-Summ} summaries carry higher semantic similarity w.r.t the complete document than the reference summaries for all four data-sets.

\begin{table}[!h]
\scriptsize
\centering
\caption{Comparison of semantic similarity w.r.t original documents for four data-sets. Competing algorithms are:   E-Summ (ES), TextRank (TR),  LexRank (LR), LSARank (LS), BANDITSUM (BS), NEUSUM (NS), REFRESH (RF). `-' indicates that summaries are not available for the data-set.}
\begin{tabular}{|l|c|c|c|c|c|c|c|}
\hline
&\multicolumn{4}{c|}{\textbf{Extractive methods}}&\multicolumn{3}{c|}{\textbf{Abstractive methods}}\\
\hline
&\textbf{ES}&\textbf{TR}&\textbf{LR}&\textbf{LS}&\textbf{BS}&\textbf{NS}&\textbf{RF}\\
\hline
DUC2001 &	\textbf{67.27}&62.08&59.76&57.70&-&-&-\\
\hline
DUC2002	&\textbf{71.17}	&65.25&63.09&61.97&-&-&-\\
\hline
CNN&\textbf{66.13}&65.75&54.79&59.66&50.96&54.49&52.59\\
\hline
DailyMail&\textbf{74.25}&70.88&62.22&67.31&50.49&50.69&62.23\\
\hline
\end{tabular}
\label{table-semantic-similarity}
\end{table}

We also compare the semantic similarity of \textit{E-Summ} summaries with three unsupervised methods and an equal number of neural methods. Unsupervised methods, TextRank  and  LexRank offer a strong baseline as evident by ROUGE results presented earlier (Tables \ref{table-duc2001-recall} - \ref{table-cnnDm-fmeasure}). LSARank\textsuperscript{\ref{footnote-miso}} \citep{steinberger2004using} is included for comparison because of its topic modelling based approach for summarization, which is similar to the approach followed by \textit{E-Summ}. Three neural methods - BANDITSUM, NEUSUM, and REFRESH, which exhibit high ROUGE scores (Table \ref{table-cnnDm-fmeasure}) are included for comparison because of availability of their CNN \& DailyMail summaries\footnote{We used BANDITSUM, NEUSUM, REFRESH summaries of CNN and DailyMail data-sets available at \url{https://github.com/Yale-LILY/SummEval} \citep{fabbri2020summeval}.}.

It is evident from Table \ref{table-semantic-similarity} that semantic similarity scores of \textit{E-Summ} summaries are  highest among extractive methods for DUC data-sets.  \textit{E-Summ} also beats all the methods for CNN and DailyMail data-sets. Evidently, \textit{E-Summ} summaries capture more representative textual information than those generated by the selected competing methods.

\subsection{Language Independence of  E-Summ}
\label{sec-language-agnostic}
In this section, we substantiate the claim of language independence of \textit{E-Summ} using the documents in two Indian languages (Hindi and Marathi) and three European languages (German, Spanish and  French). All language documents, except Hindi, are sourced from Multiling\footnote{\url{http://multiling.iit.demokritos.gr/pages/view/1571/datasets}} data-sets. Multiling is a community-driven initiative to promote NLP research in languages  other than English \citep{multiling2013overview, multiling2015overview, multiling2017overview}. 
%data-sets for multilingual automatic summarization tasks for the domains other than news articles

India is a country with a diverse set of $22$ official languages\footnote{\url{ https://www.mha.gov.in/sites/default/files/EighthSchedule_19052017.pdf}}. Most of these languages have limited language resources for performing common NLP tasks. Developing  technologies for these languages is a thrust area for Government in the country. We choose Hindi for investigating the language independence of \textit{E-Summ} because it is the most-spoken\footnote{\url{https://www.censusindia.gov.in/2011Census/C-16_25062018_NEW.pdf}, Page-6} language in India, and also the official language\footnote{\url{http://www.mea.gov.in/Images/pdf1/Part17.pdf}} alongside English. However, to the best of authors' knowledge no Hindi benchmark data-set is available for single document summarization. We also experiment with Marathi language documents, which is the official and spoken language in  Maharashtra, a state situated in the south-west part of India. Marathi language documents for single document summarization are available in Multiling 2017 data-set.
%Besides, we choose Marathi language documents available in  Multiling 2017 data-set. Marathi is another official language in India, spoken in south-west state of Maharashtra.
%
%https://www.worldatlas.com/articles/hindi-speaking-countries.html
%Multiling 2013 data-set consists of documents in $40$ languages with $30$ documents of each language for the task of single document summarization. Each document in the data-set is accompanied by a gold standard reference summary and target summary length.
 
\begin{figure}[h!]
\includegraphics[width=0.95\textwidth]{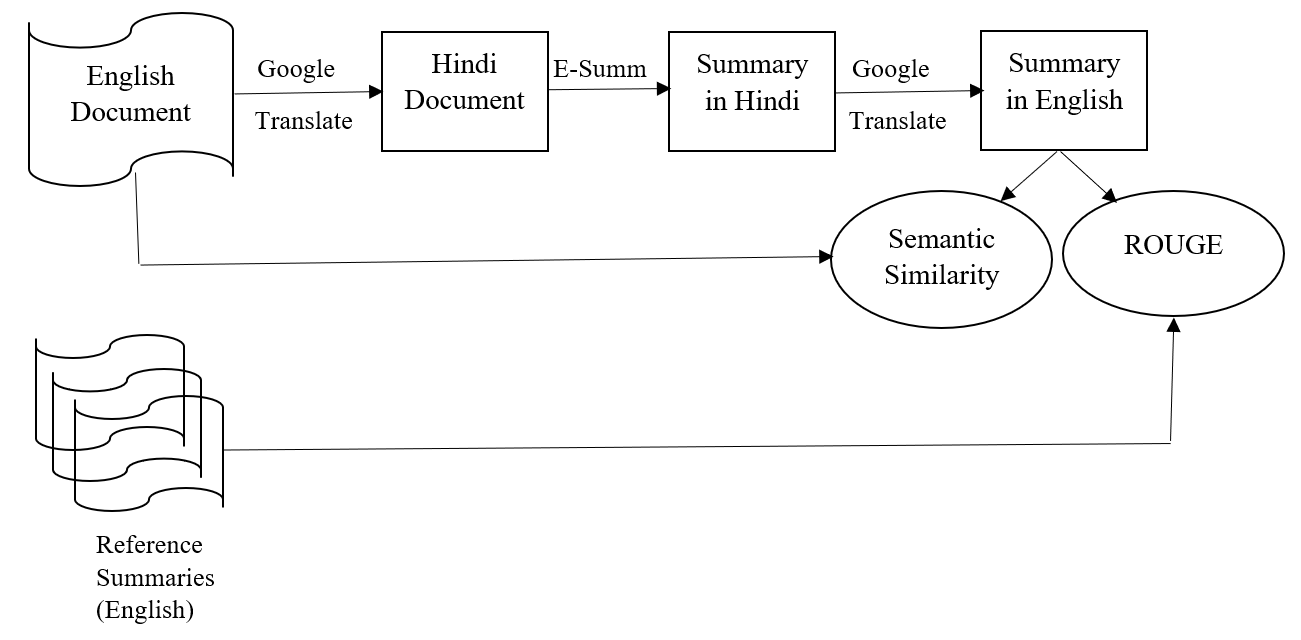}
\caption{Pipeline for creating Hindi summaries from English documents and evaluating summary quality}
\label{fig-hindi-pipeline}
\end{figure}

We employ Google translate technology to scrutinize the capability of \textit{E-Summ} for Hindi documents and present detailed results for two documents.  We first convert the English document to Hindi, sentence-by-sentence, using Google translate. Next, we tokenize sentences, filter punctuation and stop-words\footnote{We use Hindi language  stop-words list from \url{https://github.com/taranjeet/hindi-tokenizer/blob/master/stopwords.txt}  \citep{kunchukuttan2020indicnlp}. } to create a binary term-sentence incidence matrix of the translated document. We generate \textit{E-Summ} summary of the translated document and  translate the Hindi summary  back to English for evaluation using the standard ROUGE toolkit and  semantic similarity measure. Fig. \ref{fig-hindi-pipeline} describes the pipeline to assess the performance of \textit{E-Summ} algorithm for Hindi documents. This somewhat quirky sequence of steps allows summarization of an English document in any language using \textit{E-Summ} algorithm and quantitatively assess the summary quality. Admittedly, the noise introduced in two back-to-back  machine translations is expected to degrade the  resulting quality scores.

We select two documents from DUC2002 data-set (Document FBIS4-42178 and AP890606-0116), each with two reference summaries. The documents were selected because of their diverse R-L scores ($83.17$ and  $29.29$, respectively). Table \ref{table-hindi-summary-performance} presents the result  of the experiment. Rows 1 - 3 show the lengths of the generated summaries, Row 4 specifies the lengths of the two reference summaries.   Rows 5 and 6 show ROUGE scores of the Hindi summaries translated to English and \textit{E-Summ} summaries of the original documents, respectively. Rows 7 - 9 show semantic similarity with original document for Hindi translated to English, \textit{E-Summ} English, and reference summaries.

\begin{table}[h!]
\scriptsize
\centering
\caption{Performance evaluation of E-Summ summaries of two DUC2002 documents translated to Hindi following the protocol shown in Fig. \ref{fig-hindi-pipeline}. }
\begin{tabular}{|p{0.1\textwidth}|p{0.3\textwidth}|p{0.2\textwidth}|p{0.2\textwidth}|}
%\begin{tabular}{|c|c|c|}
\hline
Row &&\textbf{FBIS4-4217}8&\textbf{AP890606-0116}\\
%\multicolumn{2}{|c|}{}&FBIS4-42178&AP890606-0116\\
\hline
&\multicolumn{3}{c|}{Summary Length (in words)}\\
\hline
1&E-Summ Hindi Summary&127&104\\
\hline
2&Translated English Summary&96&90\\
\hline
3&E-Summ English Summary&123&116\\
\hline
4&Reference Summaries&99/100& 103/103\\
\hline
&\multicolumn{3}{c|}{ROUGE (R-1/ R-2/ R-L)}\\
\hline
5&Translated English Summary&72.28/39.0/71.29& 21.21/3.06/18.18\\
\hline
6&E-Summ English Summary&85.15/58.0/83.17&31.31/6.12/29.29\\
\hline
&\multicolumn{3}{c|}{Semantic Similarity with Document (in \%)}\\
\hline
7&Translated English Summary&73.87&51.22\\
\hline
8&E-Summ English Summary&93.99&63.65\\
\hline
9&Reference Summaries&66.26/77.22&31.40/31.13\\
\hline
\end{tabular}
\label{table-hindi-summary-performance}
\end{table}

As expected, the quality of translated Hindi summaries\footnote{The original documents and summaries are presented in Appendix \ref{appendix-hindi}.}  for two documents is lower than those of corresponding English summaries using both semantic similarity and ROUGE score. Two successive machine translations alter the vocabulary resulting in lower lexical overlap with reference summaries and diminish semantic similarity with the original document. Reduction in summary lengths due to translation (Rows 1 and 2) further degrades the performance compared to \textit{E-Summ} English summaries. These experiments assert competence of  \textit{E-Summ} algorithm to summarize a non-English a document with reasonable quality scores.

Multiling 2017 data-set \citep{giannakopoulos2017multiling} contains thirty Marathi language documents  with corresponding summary lengths. We pre-process these documents by tokenizing sentences\footnote{\label{hindi-marathi-tokenizer}We use  sentence tokenizer and word tokenizer implementation from \url{https://github.com/anoopkunchukuttan/indic\_nlp\_library}.}, filtering punctuation and removing stop-words\footnote{We use Marathi language  stop-words list from \url{https://github.com/stopwords-iso/stopwords-mr/blob/master/stopwords-mr.txt}.} to create the binary term-sentence incidence matrix for each document and apply \textit{E-Summ} algorithm. Since there are no gold standard summaries available, we assess the summary quality by finding the semantic similarity with the original text.  The average semantic similarity of \textit{E-Summ} summaries with the original document is found to be $89\%$. \textit{E-Summ} summaries of Marathi documents with best and least performance are presented in Appendix \ref{appendix-marathi}.

\begin{table}[!h]
\scriptsize
\centering
\caption{Performance of E-Summ on German, Spanish and French documents in Multiling 2013 data-set. $SS_{doc}$: Semantic similarity, `-': Not reported or summaries not available.}
\begin{tabular}{|l|c|c|c|c|}
\hline
&\multicolumn{4}{c|}{\textbf{GERMAN}}\\
\hline
&\textbf{ROUGE-1}&\textbf{ROUGE-2}&\textbf{ROUGE-L}&\textbf{$SS_{doc}$ }\\
\hline
E-Summ &26.79&	3.76&	\textbf{23.50}&	\textbf{41.35}\\
\hline
Multiling 2013 Best & \textbf{34.00} & 4.00 & - & - \\
\hline
TextRank &24.81	&\textbf{4.05}&	21.64&	40.40\\
\hline
Lead-3 &17.17&	2.56&	14.97&	29.12\\
\hline
&\multicolumn{4}{c|}{\textbf{SPANISH}}\\
\hline
&\textbf{ROUGE-1}&\textbf{ROUGE-2}&\textbf{ROUGE-L}&\textbf{$SS_{doc}$ }\\
\hline
E-Summ &\textbf{42.47}&	\textbf{9.35}&	\textbf{37.81}&	\textbf{61.86}\\
\hline
TextRank &39.70	&8.89&	35.14&	59.00\\
\hline
Lead-3 &17.12&	4.72&	15.64&	39.43\\
\hline
&\multicolumn{4}{c|}{\textbf{FRENCH}}\\
\hline
&\textbf{ROUGE-1}&\textbf{ROUGE-2}&\textbf{ROUGE-L}&\textbf{$SS_{doc}$ }\\
\hline
E-Summ &\textbf{41.26}&	\textbf{9.03}&	\textbf{35.60}&	\textbf{58.92}\\
\hline
TextRank &37.12	&8.56&	32.57&	56.36\\
\hline
Lead-3 &17.40&	4.43&	15.27&	37.15\\
\hline
\end{tabular}
\label{table-global-languages}
\end{table}

We further experiment with three European languages - German, Spanish and French to  scrutinize the language-independence feature of \textit{E-Summ} algorithm. We consider the test set documents in Multiling 2013 data-set for the three languages, each having thirty documents along with the gold standard reference summaries. We compare the ROUGE and semantic similarity of \textit{E-Summ} summaries with those of TextRank and  Lead-3, both of which offer strong baseline performances. Following the ROUGE evaluation criteria used in Multiling 2013 shared task (HSS scoring), we truncate the length of system summaries to the length of gold standard reference summaries. Of the three European languages used for our experiments, the participant systems of Multiling 2013 report results only for German language documents. Accordingly, we compare the performance of German test set documents with the winning ROUGE scores of Multiling 2013  \citep{multiling2013overview}.  Due to the non-availability of results and summaries for Spanish and French documents, we compare the ROUGE performance with TextRank and Lead-3.

Table \ref{table-global-languages} presents the comparative performance evaluation of \textit{E-Summ} algorithm for the three languages. We observe that \textit{E-Summ}  exhibits weak performance compared to the winners of Multiling 2013 for German (System AIC for ROUGE-1 in Fig. 1 and System MD1 for ROUGE-2 in Fig. 2 of \cite{multiling2013overview}). However, ROUGE performance of \textit{E-Summ} is better than TextRank and Lead-3 algorithms for the three languages, except for slight degradation in ROUGE-2 score for German language documents.   We also note higher values of semantic similarity scores  of \textit{E-Summ} summaries compared to TextRank and Lead-3 summaries.
%%%%%%%%%%%%%%%%%%%%%%%%%%%%%%%%%%%%%%%%%%%%%%%%%%%%%
%%%%%%%%%%    DOMAIN INDEPENDENCE   %%%%%%%%%%%%%%%%%
%%%%%%%%%%%%%%%%%%%%%%%%%%%%%%%%%%%%%%%%%%%%%%%%%%%%%

\subsection{Domain Independence of E-Summ}
\label{sec-scientific-doc-summarization}
We summarize science articles and generic documents  to substantiate the claim of domain independence of \textit{E-Summ} algorithm.

We use articles from CL-SciSumm Shared Task data-sets 2016 - 2020 \citep{clscisummtask-2016, scisumm2017overview, scisumm2018overview, scisumm2019overview, scisumm2020overview}, each split into training, development, and test sets.  Each article is accompanied by three types of summaries - (i) abstract, written by the author of the paper, (ii) community summary, created using citation spans of the paper, and (iii) human-written summaries by the annotators. We evaluate the performance of \textit{E-Summ} on the test sets w.r.t human written reference summaries of these documents. Based on the earlier empirical observation  that \textit{Abstract, Introduction}, and \textit{Conclusion} sections of the scientific articles are germane for summary \citep{kavila2015extractive, cachola2020tldr}, we selectively target these sections for  generic, concise, and  informative sentences to be included in summaries. 

Columns 2 and 3 in Table \ref{table-clscisumm-results} present   2-F and SU4-F ROUGE scores of \textit{E-Summ} summaries against the published scores of the best algorithm for each year. We report SOTA performance \citep{yasunaga2019scisummnet}  on CL-SciSumm 2016 test set. Despite the absence of training and extraneous knowledge, \textit{E-Summ} performs better than the base model trained over thirty documents. However, the performance of the model trained over $1000$ documents is significantly superior, confirming the importance of large training sets for high quality performance of neural summarization methods. 

\begin{table}[h!]
\scriptsize
\centering
\caption{Performance comparison of E-Summ for CL-SciSumm 2016-2020 test sets. 2-F: F-score ROUGE-2, SU4-F: F-score ROUGE-SU4.  Best performance is shown in boldface. `-': score not reported, * base model trained on 30 documents, ** best model trained on 1000 documents.}
\begin{tabular}{|l|c|c|}
\hline
\textbf{CL-SciSumm 2016}&\textbf{2-F}&\textbf{SU4-F}\\
\hline
E-Summ&24.11&16.32\\    
\hline
\cite{yasunaga2019scisummnet}*&18.46&12.21\\
\hline
\cite{yasunaga2019scisummnet}**&\textbf{31.54} &\textbf{24.36}\\
\hline
\textbf{CL-SciSumm 2017}& \multicolumn{2}{c|}{}\\
\hline
E-Summ&25.45&\textbf{18.60}\\    
\hline
CIST \citep{best2017cist, scisumm2017overview}  &\textbf{27.50}&17.80\\
\hline
& \multicolumn{2}{c|}{Test sets for 2018-2020 are same.}\\
\hline
\textbf{CL-SciSumm 2018}& \multicolumn{2}{c|}{}\\
\hline
E-Summ&25.53&	20.75\\    
\hline
LaSTUS/TALN+INCO \citep{best2018lastus, scisumm2018overview}  &\textbf{28.80}&\textbf{24.00}\\
\hline
\textbf{CL-SciSumm 2019}& \multicolumn{2}{c|}{}\\
\hline
CIST \citep{best2019cist, scisumm2019overview}&27.80&20.00\\
\hline
\textbf{CL-SciSumm 2020}& \multicolumn{2}{c|}{}\\
\hline
AUTH \citep{scisumm2020overview}& 22.00&- \\
\hline
\end{tabular}
\label{table-clscisumm-results}
\end{table}
For CL-SciSumm 2017 data-set, performance of \textit{E-Summ} is lower than the best scoring system proposed by \citet{best2017cist}. The winning method employs statistical features, estimate feature weights and ensure non-redundancy using Determinantal Point Processes (DPPs) based sentence sampling to select sentences for summary. The test sets for CL-SciSumm 2018 - 2020 are same, and hence there is one row for \textit{E-Summ} performance. Winning system for CLSciSumm 2018 \citep{best2018lastus} beats \textit{E-Summ}. The method employs convolutional neural network to learn relation between context based document features, and uses likelihood based scoring function. The dip in the performance of winning systems for CL-SciSumm 2019 and CL-SciSumm 2020 is unexpected. \citet{best2019cist} employ statistical feature model and neural language model to extract summary sentences using DPP sampling. \citet{best2020auth} follow an abstractive summarization technique using PEGASUS model pre-trained on the arXiv data-set and generate the summary of scientific article based on abstract and cited text spans.

Admittedly, summaries generated by the  best model proposed by \citep{best2017cist, best2018lastus, best2019cist, yasunaga2019scisummnet} are significantly better than those produced by \textit{E-Summ}. However, frugality in terms of  human curated knowledge makes \textit{E-Summ} particularly attractive for summarizing new research articles with little or no citation information available.

WikiHow is a large scale data-set consisting of article-summary pairs prepared from an online knowledge base, written by different human authors and covering a wide range of topics with diverse writing styles  \citep{koupaee2018wikihow}. The data-set is divided into training, validation and testing sets consisting of $1,68,126 / 6,000 / 6,000$ documents respectively. Following previous works \citep{zhong2020extractive, dou2021gsum}, we extract four sentences for \textit{E-Summ} summaries of WikiHow documents.

\begin{table}[!h]
\scriptsize
\centering
\caption{Performance comparison of the proposed method for WikiHow test set based on ROUGE-F scores. Best performance is in boldface. E: Extractive methods, A: Abstractive methods.}
\begin{tabular}{|l|l|c|c|c|}
\hline
Type&&\textbf{ROUGE-1} & \textbf{ROUGE-2} & \textbf{ROUGE-L}\\
\hline
\multirow{5}{0.8em}{E}
&E-Summ	&25.72&	6.44&	23.63\\
%&BertExt \citep{liu2019text}&	30.40 &8.67 &28.32\\
&MatchSum \citep{zhong2020extractive}& 31.85 & 8.98 & 29.58\\
&CUPS \citep{desai2020compressive}&30.94 &9.06& 28.81\\
&LFIP-SUM \citep{wikihow_extractive_jang2021}&24.28&5.32&18.69\\
\hline
\multirow{2}{0.8em}{A} 
&BART \citep{lewis2019bart}&41.46& \textbf{17.80}& 39.89\\
&GSum \citep{dou2021gsum} &\textbf{41.74} &17.73& \textbf{40.09}\\
\hline
\end{tabular}
\label{table-wikihow}
\end{table}

Table \ref{table-wikihow} presents comparative evaluation of \textit{E-Summ} with recent  methods evaluated on WikiHow data-set.  BART and GSum exhibit  SOTA  performance on this data-set.  Bert based extractive neural summarization methods \citep{zhong2020extractive, desai2020compressive}  lose by  a wide margin of $\approx$ $10\%$. LFIP-SUM and \textit{E-Summ} have comparable performance, which is clearly weak. LFIP-SUM is an unsupervised extractive summarization method, which formulates summarization as an integer linear programming problem based on pre-trained sentence embeddings and uses Principal Component Analysis for sentence importance and extraction.

Since, WikiHow summaries are highly abstractive   with average compression ratio of $2.38\%$ \citep{koupaee2018wikihow}, abstractive summarization methods are expected  to score  much better than extractive  methods. High extent of  abstraction and lexical variability in  WikiHow gold standard reference summaries  inevitably lower ROUGE performance of extractive summarization methods as revealed by our investigation.  

%%%%%%%%%%%%%%%%%%%%%%%%%%%%%%%%%%%%%%%%%%%
%%%%%%%%%%  SECTION - 8 %%%%%%%%%%%%%%%%%%%
%%%%%%%%%%%%%%%%%%%%%%%%%%%%%%%%%%%%%%%%%%%

\section{Discussion}
\label{sec-discussion}
This section presents an insightful analysis of the consolidated results presented in the previous section (Tables \ref{table-duc2001-recall} - \ref{table-wikihow}).

\noindent
\textbf{State-of-the-art in  Document Summarization}\\
The vast majority of extractive, non-neural    summarization algorithms use four data-sets for performance evaluation,  exhibiting an interesting pattern.  Unsupervised   summarization methods majorly evaluate performance on  DUC data-sets (\cite{fang2017wordcorank, alguliyev2019cosum, saini2019extractive} among other recent works), while deep neural  summarization methods use CNN and  DailyMail data-sets. However, some neural methods train on CNN and  DailyMail data-sets and test performance on  DUC2002 considering it as out-of-domain data-set  (\cite{cheng2016neural, nallapati2017summarunner, zhou2018neural, dong2018banditsum, al2018hierarchical}).  Recently added NEWSROOM data-set \citep{grusky2018newsroom} consisting of news articles and corresponding  human written reference summaries are gradually gaining popularity among summarization research community.

Table \ref{table-conclusion} lists performance scores of \textit{E-Summ} algorithm and the winner for each  data-set, thereby recording the gap between \textit{E-Summ} and best performance in terms of ROUGE scores.  Entries marked `-' indicate missing evaluation of the algorithm, and `*' indicates that the algorithm is not the top-scorer for the corresponding data-set.

\begin{table}[!h]
\scriptsize
\centering
\caption{Consolidated results for all four data-sets. R-1: ROUGE-1, R-2: ROUGE-2, R-L: ROUGE-L. * denotes that algorithm is not top-scorer. -: Not Reported.}
\begin{tabular}{|l|c|c|c|c|c|c|c|c|c|}
\hline
&\multicolumn{3}{c|}{\textbf{DUC2001}} & \multicolumn{3}{c|}{\textbf{DUC2002}}& \multicolumn{3}{c|}{\textbf{CNN \& DailyMail}}\\
\hline
&R-1&R-2&R-L&R-1&R-2&R-L&R-1&R-2&R-L\\
\hline
E-Summ&45.6&15.71&40.22&50.39&21.16&45.19&30.97&10.94&27.78\\    
\hline
\hline
CoRank+ (2017) &-&-&-&52.6&*&*&-&-&-\\
\hline
\cite{saini2019extractive}&50.24&29.24&-&*&28.85&-&-&-&-\\
\hline
\hline
HSSAS (2018) &-&-&-&*&*&48.8&*&*&*\\
\hline
GSum (2021)&-&-&-&-&-&-&45.94& 22.32 &42.48\\
\hline
\end{tabular}
\label{table-conclusion}
\end{table}

GSum \citep{dou2021gsum} delivers state-of-the-art performance (all three ROUGE scores) for one data-set. The method uses pre-trained BART with additional guidance as input to control the output. Notably, the selective presentation of evaluation results for   data-sets makes fair comparison of algorithms difficult. Under the present circumstances, a method can at best be designated \textit{state-of-the-art} for a \textit{specific data-set}. It is reasonable to conclude that state-of-the-art for generic extractive summarization is yet to be achieved.

\textit{E-Summ} algorithm, which is well-grounded in information theory, generates summaries with high semantic similarities, even though it suffers from  relatively lower lexical matching measured by ROUGE metric.  We believe that information theoretic methods like \textit{E-Summ} have a high potential to evolve and occupy space in the bouquet of summarization algorithms. 

\noindent
\textbf{Deep Neural Vs. Non-neural methods}\\
Experiments reported in Sec. \ref{sec-experimental-results} show that  ROUGE scores of  summaries generated by deep neural methods are generally  higher for all data-sets, suggesting that these methods hold more promise than their unsupervised counterparts.

Understandably, the  advantage comes with a price of long model training time and training-data preparation time for neural methods. Dependence on language models, domain- and collection dependence, lack of transparency, and interpretability are ancillary costs of these methods. Furthermore, high ROUGE score performance exhibited by recent neural methods  for \textit{specific data-sets} gives an inkling of weak summarization abilities of these methods akin to weak AI.

Though neural methods easily beat \textit{E-Summ} algorithm, \textit{E-Summ} establishes the promise of information theoretic approach for unsupervised extractive document summarization. Vanilla \textit{E-Summ} can be bolstered by effective sentence selection methods to tear apart distracting topics and reveal semantically most relevant sentences. Some interesting ideas that can fortify \textit{E-Summ} include DPP for sentence selection \citep{best2017cist, best2019cist}, graph-based methods for sentence selection \citep{sentencecentralityzheng2019, gupta2019entailment}, multi-objective optimization for  sentence selection \citep{saini2019extractive, mishra2021scientific}, BART for abstraction \citep{chaturvedi2020divide, dou2021gsum}, using embedding based similarity for reducing redundancy \citep{hailu2020framework, zhong2020extractive} etc.

\noindent
\textbf{Comparison of Document Summarization Applications }\\
Applications of text summarization  have been rising monotonically and will continue to do so in the foreseeable future. There exists a considerable demand for summarization tools in diverse \textit{domains} and \textit{genres}, and more so in the context of \textit{purpose} (\cite{kanapala2019text}). Existing summarization methods do not take cognizance of either genre or purpose, thereby missing salient cues exposed by structure, writing style, vocabulary, etc. They also are oblivious to the requirement of different types of summaries based  on evolving user’s needs (\cite{lloret2012textsurvey}).   

Most existing deep neural summarization methods are trained on documents belonging to the genre ``news" and  their performance over scientific articles, literary documents, blogs and web pages, reports, letters and memos, etc. is yet to be examined. However, once trained appropriately, deep neural methods hold high promise for high precision summarization of scholarly documents in science, technology, social science, law and international relations, etc. Unsupervised summarization methods like COSUM, CoRank+, \textit{E-Summ} etc. are independent of collection,  with the design inspired by intuitive ideas for summarization by humans, and backed by sound computational techniques. Consequently, they are expected to exhibit more predictable  performance across genres.

On-the-fly  summarization for web documents is one of the  most desired text analytics tasks in the current era. With deeper penetration of digital services, improving literacy rates, advancing language technology, more people are accessing the internet to stay connected via online social networks, to communicate in  office and personal domains, satisfy their knowledge needs, etc. Summarization integrated with browsers may become as commonplace as language translation tools over time. To meet this requirement, it is pragmatic to develop  language agnostic, lean, and fast methods capable of summarizing generic documents. Recently, Dhaliwal et al. proposed a device based model that employs character-level neural architecture for extractive text summarization \citep{on-device-summarization}. So far, real time on-device summarization is an un-chartered territory and we envisage rapid  developments in this direction.

\noindent
\textbf{Personalized Summary}\\
A summary is \textit{informative} for a user if it adds to her personal  knowledge. Peyrard claims that once the reader's background knowledge is modeled, it can be blended with an information-theoretic framework to generate informative personalized summaries \citep{peyrard2019simple}.  Generating user-centric summaries entails capturing user background knowledge ($\mathbf K$) in terms of semantic units and  identifying those that  maximize \textit{relevance} and minimize \textit{redundancy}.  Minimizing  KL-Divergence between the distribution of the semantics units in the summary (\textbf{S}) and original document (\textbf{D}) addresses \textit{relevance} and  \textit{redundancy},  while the amount of new information contained in a summary is given by the cross-entropy between $\mathbf K$ and \textbf{S}. This is a promising line of research and requires addressing  additional challenges in the area human-computer interactions.

\noindent
\textbf{Limitations of E-Summ}\\
\textit{E-Summ} is an efficient algorithm to generate extractive summaries without any dependence on extraneous knowledge and with no training, tuning and feature selection  overheads. Besides, the method holds the promise of domain and language independence.

\begin{table}[!h]
\scriptsize
\centering
\caption{Execution time for twelve documents from DUC2002 data-set. NMF: time for matrix factorization, Knapsack: time for sentence selection using Knapsack algorithm, Total: total time for execution of E-Summ algorithm. m: number of terms, n: number of sentences in the document. Reported times are in seconds. }
\begin{tabular}{|l|c|l|c|l|c|l|}
\hline
& \multicolumn{3}{c|}{\textbf{Document Statistics}}  & \multicolumn{3}{c|}{\textbf{Execution Time}} \\ \hline
& Size (KB) &  A (m $\times$ n)   & \#topics & NMF     & Knapsack   & Total   \\ 
\hline
LA080890-0078  & 1  & (106 $\times$ 12)  & 2& 5.490E-03	&1.128E-04&	1.894E-02\\ 
\hline
AP900807-0029 & 2  & (183 $\times$ 23)  & 2 & 5.295E-03&1.066E-04&	3.063E-02 \\ 
\hline
WSJ880412-0015  & 3 & (258 $\times$ 28)  & 6  & 1.309E-02&1.150E-04&	4.013E-02 \\ 
\hline
WSJ870122-0100 & 4  & (288 $\times$ 33)  & 5  & 1.189E-02&	6.309E-05&	4.724E-02 \\ 
\hline
LA102389-0075& 5 & (342 $\times$ 53)  & 9 & 2.086E-02&	3.768E-04&	7.623E-02 \\ 
\hline
WSJ891019-0086 & 6  & (375 $\times$ 42)  & 7 & 2.147E-02&	1.140E-04&	6.776E-02 \\ 
\hline
FT943-3897  & 7  & (493 $\times$ 56)  & 6 & 1.253E-02&	9.730E-05&	7.727E-02 \\
\hline
LA012090-0110 & 8 & (410 $\times$ 66)  & 7 & 2.448E-02&	1.380E-04&	9.202E-02 \\ 
\hline
WSJ881004-0111  & 10 & (649 $\times$ 64)  & 2 & 1.096E-02&	4.385E-05&	9.746E-02 \\ 
\hline
LA101590-0066 & 16 & (911 $\times$ 147) & 10 & 5.808E-02&	4.489E-04&	2.948E-01 \\ 
\hline
LA042789-0025& 17 & (984 $\times$ 188) & 14 & 5.122E-02&9.550E-04&	3.473E-01 \\ 
\hline
LA011990-0091& 20& (987 $\times$ 173) & 11 & 5.289E-02&	5.216E-04&	3.399E-01 \\ 
\hline
\end{tabular}
\label{table-nmf-timings}
\end{table}
Intuitively, a long document would generate a high dimensional term-sentence matrix and a higher number of latent topics, raising the memory requirement and computational expense of NMF. To investigate the scalability of \textit{E-Summ}, we select twelve documents of different lengths from DUC2002 data-set, and note the size of term-sentence matrix, number of topics and break-up of  time for factorization and sentence selection  (Table \ref{table-nmf-timings}). It is evident that dimensions of term-sentence matrix and number of latent topics increase with size of the document, as expected. However, the computational expense of NMF does not increase significantly with the size of the input matrix, thanks to efficient NMF solvers. Since the size of summary is small and same (100 words) for all documents, execution time for Knapsack algorithm is negligible and nearly same for all documents.  However,  \textit{E-Summ} algorithm suffers from following two limitations. 
\begin{enumerate}[(i)]
    \item Long Summaries: \textit{E-Summ} is not suitable for generating long summaries. Extracting a long summary requires solution for large size knapsack resulting in high execution cost of Knapsack algorithm. We  generate summaries of varying lengths  for  the longest document (LA011990-0091 - 3181 words) in DUC2002 data-set, and observe the timing for execution of  Knapsack algorithm. Table \ref{table-running-time}  reveals that if long summaries are required than sentence selection by  Knapsack algorithm is the bottleneck. 
   \begin{table}[!h]
        \scriptsize
        \centering
        \caption{Execution time of Knapsack algorithm (in seconds) for  document no. LA011990-0091 in DUC2002 data-set for different summary lengths (L).}
        \begin{tabular}{|l|c|c|c|c|c|c|c|}
        \hline
        L (words) & \textbf{100} & \textbf{200} & \textbf{300} & \textbf{400} & \textbf{500} & \textbf{600} & \textbf{700}\\
        \hline
        Knapsack&5.216E-04&1.091E-03&1.309E-03&	1.268E-03&	1.637E-01	&4.050E+00&	1.585E+02\\
        %Knapsack & 0.000 & 0.001 & 0.003 & 0.007 & 0.153 & 4.388 & 172.254\\
        \hline
        Total &3.399E-01&3.509E-01&	3.456E-01&	3.469E-01&	6.860E-01&	4.680E+00&	1.592E+02\\
        %Total&	0.333&	0.343&0.365&0.529&	0.710&	5.071&	173.097\\
        \hline
        \end{tabular}
        \label{table-running-time}
    \end{table}
    
    \item Highlight Extraction: \textit{E-Summ} is not suitable for generating highlights or TLDR summaries. This is evident from the weak results on WikiHow and CNN/Daily mail data-sets, both of  which have abstracted highlights of the articles. Scope of summarization methods is subtly different from  that of highlights generation task. While both are expected to have document wide coverage, summary of an article is expected to be smooth and coherent, whereas highlights are short sentences that provide a focused  overview of the article.  The two have been recognized as  two independent tasks with contrasting objectives, particularly in the context of scientific articles \citep{cagliero2020extractinghighlights}. Extractive summarization methods are not suitable for \textit{highlight} generation, and need to be blended with abstractive summarization models.
\end{enumerate}

%%%%%%%%%%%%%%%%%%%%%%%%%%%%%%%%%%%%%%%%%%
%%%%%%%%%%  SECTION - 9 %%%%%%%%%%%%%%%%%%%
%%%%%%%%%%%%%%%%%%%%%%%%%%%%%%%%%%%%%%%%%%%

\section{Conclusion}
\label{sec-conclusion-future-work}
In this paper, we propose an information theoretic approach for unsupervised, extractive single document summarization. We decompose the binary term-sentence matrix for a document using non-negative matrix factorization and adopt probabilistic perspective of the resulting factor matrices. We extract probability distributions of sentences and topics in latent space, the principal semantic units constituting the document. We then  leverage entropy of topics and sentences for generation of the document summary. The proposed \textit{E-Summ} algorithm  is domain-, collection-independent and is agnostic to the language of the document. Moreover, the method is explainable and fast enough to meet real-time requirements for on-the-fly summarization of web documents in  languages other than English. 

We present a comprehensive performance analysis of the proposed method on four well-known public data-sets and  present the results dataset-wise. The reported experiments reveal that the combination of NMF and information theory begets advantages of speed and  transparency but falls short of comparative performance measured by ROUGE score. Since all gold-standard summaries are abstractive (human generated),  ROUGE score measurement for extractive summary has assurance deficit. We proceed to measure semantic similarity of \textit{E-Summ} summary with the complete document  and observe that it is higher than that of the reference summaries for all data-sets.

Encouraged by our results in Sec. \ref{sec-scientific-doc-summarization}, we present \textit{E-Summ} summaries of 150\footnote{Summary length of both the sections is slightly long because the last sentence selected for summary overshoots the summary length. } words for two key sections of this article in Fig. \ref{fig-document-summary}. We  remove mathematical equations, figures and tables, and determine the number of latent topics using community detection  method (Sec. \ref{sec-topics}). Sentences, in summary, are presented in the order in which they appear in the article. We leave it to the reader to judge the quality of summary.

\begin{figure}[htbp!]
\centering
\caption{E-Summ summary of Sections \ref{sec-introduction} and \ref{sec-algorithm} of the present article}
\label{fig-document-summary}
\begin{subfigure}[b]{\textwidth}
\caption{173 words summary of Section \ref{sec-introduction} (Introduction) : Semantic similarity - 78.41}
\label{fig-sec1-summary}
\noindent\fbox{%
    \parbox{\textwidth}{%
\scriptsize{\setstretch{1.0}{
Exponential growth in textual information available online has spurred the need for automatic processing of text for various tasks that humanity performs on computing devices.\\
Proposed more than six decades ago by Luhn, modest progress in the area of automatic summarization is evident by moderate scores achieved by sophisticated deep neural methods.\\
The author further argues that community has exerted over crafting algorithms to improve performance evaluation scores on the benchmark data sets, thereby limiting progress in the science of automatic extractive document summarization.\\
Peyrard contends that the notion of importance unifies non redundancy, relevance and informativeness, the summary attributes hitherto addressed by the research community in an adhoc manner.\\
Admitting sentences and topics as crucial semantics units for document summarization, we present an in depth analysis of sentence and topic entropy in latent semantic space revealed by Non negative Matrix Factorization.\\
We also evaluate algorithmic summary quality by computing its semantic similarity with the complete document and show that the reference summaries have relatively less semantic overlap compared to E-Summ summaries.
}
}}
}
\end{subfigure}

\begin{subfigure}[b]{\textwidth}
\caption{$185$ words summary of Section \ref{sec-algorithm} (Generating Informative Summary) : Semantic similarity - 88.10}
\label{fig-sec2-summary}
\noindent\fbox{%
    \parbox{\textwidth}{%
\scriptsize{
In this section, we propose an unsupervised algorithm called E-Summ, which is an information theoretic method for extractive single document summarization.\\
Based on the ground rule that a sentence with low self information is the good representative of a topic, E-Summ identifies candidates by choosing sentences with the least self information from each latent topic and assigns a score as the sum of sentence entropy in topic space and topic entropy in sentence space as follows.\\
Considering sentence length as item weight, score as item value, and required summary length as the capacity, the Knapsack algorithm selects sentences from the set of candidates to maximize the total score for the required summary length.\\
Subsequently, in step 11, the Knapsack optimization algorithm is applied to maximize information content of the summary while
 delimiting the summary length.\\
However, when desired summary length is specified as number of sentences, application of Knapsack algorithm is omitted and top scoring candidates are selected for inclusion in summary.\\
Though NMF is an NP hard problem  , efficient  polynomial time algorithms that use local search heuristic are commonly available in libraries.
}
}}
\end{subfigure}
\end{figure}

\section{Acknowledgement}
We sincerely thank the anonymous reviewers for their valuable and constructive feedback, which has led to substantial improvement in the quality of paper. We also thank the authors of winning teams of CL-SciSumm Shared task 2018 and 2019 for sharing the summaries.

%%%%%%%%%%%%%%%%%%%%%%%%%%%%%%%%%%%%%%%%%%%
%%%%%%%%%%  SECTION - 10 %%%%%%%%%%%%%%%%%%%
%%%%%%%%%%%%%%%%%%%%%%%%%%%%%%%%%%%%%%%%%%%
\section{References}
\bibliographystyle{model5-names}\biboptions{authoryear}
\bibliography{entropy}

\newpage
% \appendix
\begin{appendices}
\section{Appendix}
\label{sec-appendix}
\subsection{Summaries of Hindi documents}
\label{appendix-hindi}

We report  \textit{E-Summ} summaries of the two documents referred in Sec. \ref{sec-language-agnostic}. For each document,  \textit{E-Summ} Hindi summary appears first, followed by its   English translation  and \textit{E-Summ} summary of the original English document.
%\textcolor{blue}{For the best scoring document, four sentences obtained in Google translation of Hindi summary are present in  English  \textit{E-Summ} summary. For the low scoring document, only one sentence is common between Hindi summary translated into English and the original English summary.}

\begin{figure}[h]
\centering
\includegraphics[width=\textwidth]{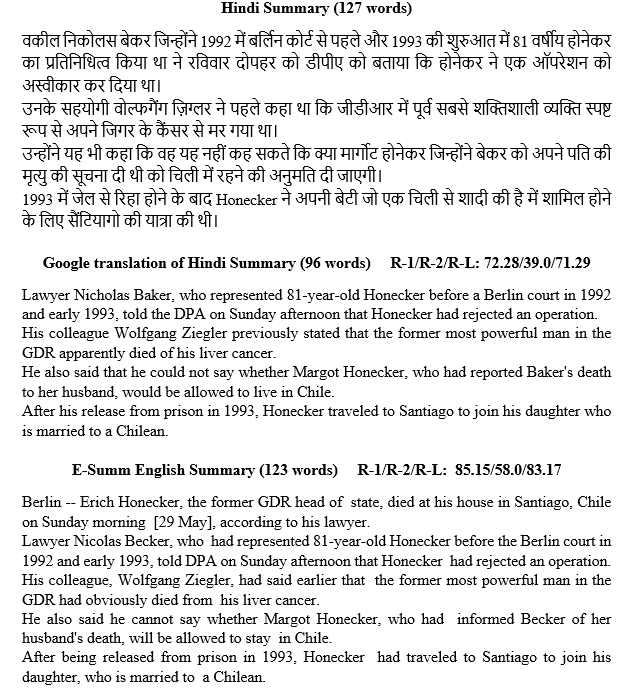}
\caption{Summaries of document no. FBIS4-42178 from DUC2002 data-set with high ROUGE score. Four sentences obtained in Google translation of Hindi summary are present in  the   \textit{E-Summ} summary of the original English document. ROUGE score of the summary translated from Hindi are lower.}
\label{fig-doc1-summaries}
\end{figure}

\begin{figure}[h!]
\centering
\includegraphics[width=\textwidth]{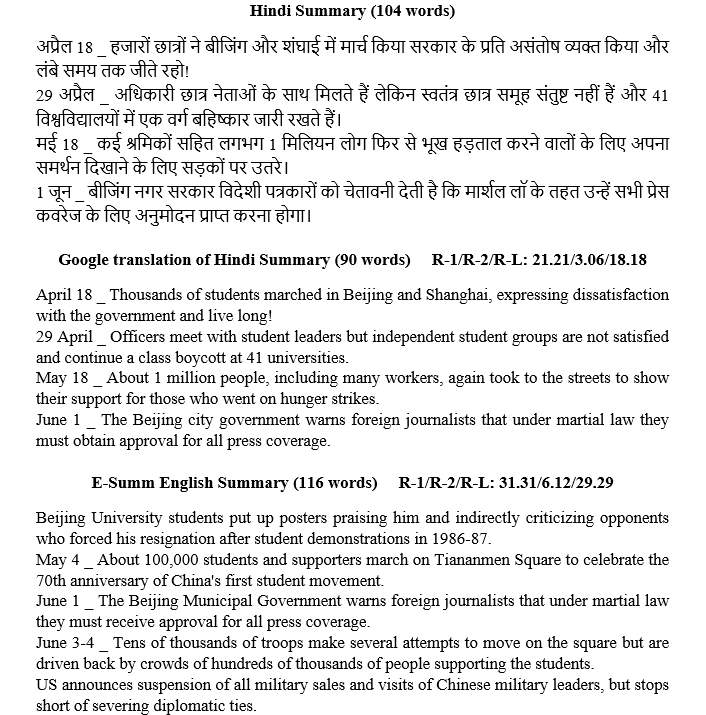}
\caption{Summaries of document no. AP890606-0116 from DUC2002 data-set  with low ROUGE score. Only one sentence obtained in Google translation of Hindi summary is present in   the   \textit{E-Summ} summary of the original English document. ROUGE score of the summary translated from Hindi are lower. }
\label{fig-doc2-summaries}
\end{figure}

\subsection{Summaries of Marathi documents}
\label{appendix-marathi}
We report \textit{E-Summ} summaries of two Marathi language documents with best and worst semantic similarity scores with their corresponding original document. 
\begin{figure}[h!]
\centering
\includegraphics[width=\textwidth]{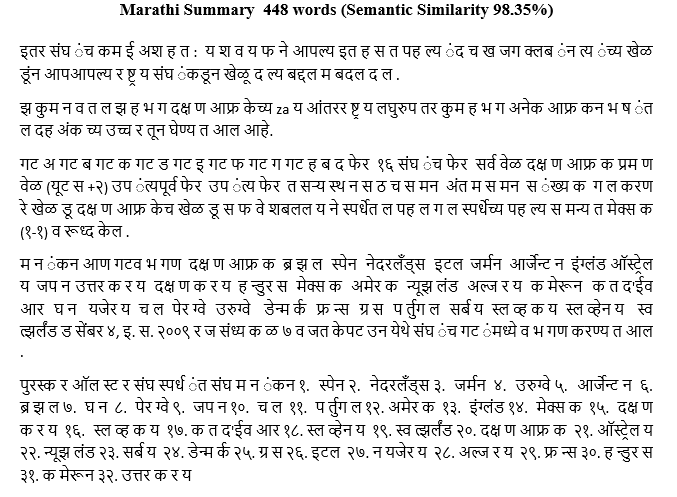}
\caption{Summary of Marathi document no. c7253fca8da3ef6a43934d0728abe0e5 from Multiling2017 data-set with highest semantic similarity.}
\label{fig-doc3-summaries}
\end{figure}

\begin{figure}[h!]
\centering
\includegraphics[width=\textwidth]{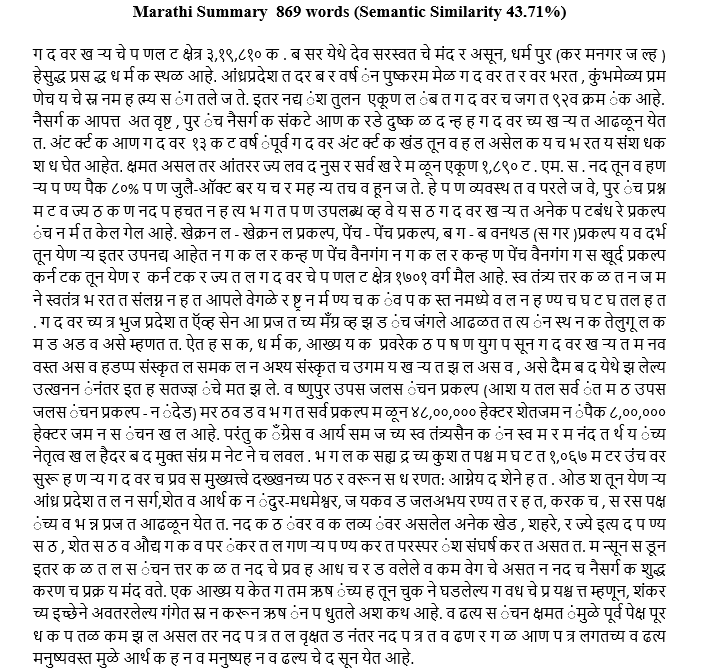}
\caption{Summary of Marathi document no. 3abd7d9dd7ff3454f3b9acd68022cc96 from Multiling2017 data-set with least semantic similarity.}
\label{fig-doc4-summaries}
\end{figure}
\end{appendices}
\end{document}